\documentclass[%
 reprint,
 amsmath,amssymb,
 aps,
]{revtex4-2}

\usepackage[colorlinks=true,linkcolor=blue, citecolor=blue]{hyperref}
\usepackage{graphicx}
\usepackage{subfigure}
\usepackage{dcolumn}
\usepackage{bm}
\usepackage[toc]{appendix}
\usepackage{titlesec}

\begin{document}

\preprint{APS/123-QED}

\title{Nuclear cluster structure effect in $^{16}$O+$^{16}$O collisions at the top RHIC energy}

\author{Xin-Li Zhao\textsuperscript{1,2,3}}
\email{zhaoxinli@usst.edu.cn}
\author{Zi-Wei Lin\textsuperscript{4}}
\email{linz@ecu.edu}
\author{You Zhou\textsuperscript{5}}
\email{you.zhou@cern.ch}
\author{Chao Zhang\textsuperscript{6}}
\email{chaoz@whut.edu.cn}
\author{Guo-Liang Ma\textsuperscript{2,3}}
\email{glma@fudan.edu.cn}

\affiliation{\textsuperscript{1}College of Science, University of Shanghai for Science and Technology, Shanghai 200093, China}
\affiliation{\textsuperscript{2}Key Laboratory of Nuclear Physics and Ion-beam Application (MOE), Institute of Modern Physics, Fudan University, Shanghai 200433, China}
\affiliation{\textsuperscript{3}Shanghai Research Center for Theoretical Nuclear Physics, NSFC and Fudan University, Shanghai 200438, China}
\affiliation{\textsuperscript{4}Department of Physics, East Carolina University, Greenville, NC 27858, USA}
\affiliation{\textsuperscript{5}Niels Bohr Institute, Jagtvej 155A, 2200 Copenhagen, Denmark}
\affiliation{\textsuperscript{6}Department of Physical Science and Technology, Wuhan University of Technology, Wuhan 430070, China}


\begin{abstract}
Using the improved string-melting version of a Multi-Phase Transport model, we investigated the impact of nuclear geometry of $^{16}$O on anisotropic flows in O+O collisions at $\sqrt{s_{_{\rm NN}}}=200$~GeV. To evaluate the influence of nuclear structure and potential alpha clustering, we implemented four candidate configurations: Woods–Saxon, tetrahedron, square, and Nuclear Lattice Effective Field Theory. Initial-state geometry is quantified via the eccentricity cumulant ratio $\varepsilon_{2}\{4\}/\varepsilon_{2}\{2\}$, which provides a robust and evolution-independent measure sensitive to configuration differences. The model reproduces $v_{2}(p_{\rm T})$ at low $p_{\rm T}$ and $v_{3}(p_{\rm T})$ across the full $p_{\rm T}$ range, with integrated $v_{2}\{2\}$ and $v_{3}\{2\}$ matching the STAR data, demonstrating that transport dynamics captures the essential collectivity in this intermediate-size system. These findings establish a baseline for extending nuclear-structure studies in O+O collisions to other energies and differential observables within a unified transport model framework.

\end{abstract}

\maketitle


\section{\label{sec:level1} Introduction}
High-energy heavy-ion collisions create extremely hot and dense conditions where quarks and gluons, normally confined within hadrons, can transition into a deconfined state of matter known as the quark-gluon plasma (QGP). This exotic state, believed to have existed microseconds after the Big Bang, has been extensively studied at the Relativistic Heavy Ion Collider (RHIC) and the Large Hadron Collider (LHC). One of the key signatures of QGP formation is anisotropic flow, which arises from the conversion of initial spatial anisotropies into final-state momentum anisotropies through the collective expansion of the system~\cite{Ollitrault:1992bk,Heinz:2013th,Zhao:2017rgg,Chen:2024aom}. In large collision systems such as Au+Au and Pb+Pb, the observed anisotropic flow $v_n$ has been successfully described by hydrodynamics models~\cite{STAR:2005gfr,Song:2017wtw,Hirano:2008hy}, indicating that QGP behaves as a nearly perfect fluid. Surprisingly, similar flow-like signals have also been observed in small systems such as $pp$ and $p$+A~\cite{ALICE:2019zfl}, sparking ongoing debates about whether QGP is formed under such systems. Due to their small size and short lifetime, it remains an open question whether hydrodynamics is applicable to these systems or whether non-equilibrium effect such as the escape mechanism is important~\cite{He:2015hfa}. As a result, the intermediate-sized systems such as O+O and Ne+Ne collisions have been proposed as ideal testbeds for systematically probing the emergence of collective behavior and its system-size dependence~\cite{Brewer:2021kiv,Huang:2023viw}. In particular, O+O collisions have attracted great attention because the oxygen nucleus, especially $^{16}$O, has long been suggested to exhibit pronounced cluster structures. This provides a unique opportunity to connect the study of anisotropic flow in relativistic collisions with long-standing questions in nuclear structure physics, thereby allowing experimental data to shed light on the role of clustering effects in high-energy heavy-ion collisions. 

A unique feature of anisotropic flow is its sensitivity to the initial spatial geometry of the colliding system, making it a potential probe of the nuclear structure~\cite{ALICE:2018lao,ALICE:2021gxt,Zhang:2021kxj,Zhang:2022fou,Lu:2023fqd}. Among light nuclei, $^{16}$O is particularly intriguing because theoretical studies predict that it may exhibit $\alpha$-clustering configurations, such as a tetrahedron or square arrangement of four $\alpha$ particles~\cite{Gamow:1930,Hoyle:1954zz,Funaki:2008gb,Ohkubo:2013rr,Funaki:2013nea}. These configurations can generate distinctive fluctuations in the initial geometry, which may manifest in final-state observables like $v_{2}$ and $v_{3}$~\cite{Broniowski:2013dia,Bozek:2014cva,Zhang:2017xda,Ma:2022dbh}. Recent theoretical studies further suggest that anisotropic flow in high-energy O+O collisions may serve as a sensitive probe of clustering in $^{16}$O~\cite{Rybczynski:2019adt,Li:2020vrg,Li:2021znq,Behera:2021zhi,Ding:2023ibq}. On the experimental side, the STAR Collaboration has recently reported the first measurements of anisotropic flow in O+O collisions at $\sqrt{s_{{\rm NN}}}=200$ GeV~\cite{Huang:2023viw}, while the LHC experiments have already released the latest results on O+O collisions at $\sqrt{s{_{\rm NN}}}=5.36$ TeV from Run 3~\cite{ALICE:2025luc,ATLAS:2025nnt,CMS:2025tga}, offering new opportunities for further exploration.

Motivated by these developments, we employ the latest version of a Multi-Phase Transport (AMPT) model~\cite{Bzdak:2014dia,Ma:2016fve,Tang:2024kot} to investigate O+O collisions. AMPT model, as a microscopic transport approach, has achieved remarkable success in describing a wide variety of observables in large collision systems such as Au+Au and Pb+Pb~\cite{Ma:2016fve,Chen:2022xpm}. These studies demonstrate that AMPT can provide valuable insight into the collective dynamics of QGP and the transformation of initial-state geometry into final-state anisotropies. However, its performance in small and intermediate systems is less satisfactory, especially for central collisions where collective behavior is expected to be most pronounced~\cite{Zhao:2021bef}. 
To address this limitation, we introduce an improved hadronization scheme in AMPT, in which the hadron formation time depends on the impact parameter. This adjustment stabilizes the parton-to-hadron transition in O+O collisions without modifying other model parameters, leading to more reasonable centrality trends for both bulk and flow observables. In this study, we focus on comparisons with STAR measurements of $v_{2}\{2\}$ and $v_{3}\{2\}$ in O+O collisions at $\sqrt{s_{{\rm NN}}}=200$~GeV. With a consistent AMPT setup that applies the same collision conditions, analysis methods, and non-flow subtraction, we compared several possible $^{16}$O configurations: Woods–Saxon (W-S), tetrahedron, square, and Nuclear Lattice Effective Field Theory (NLEFT). This allowed us to examine the effects of nuclear geometry and potential $\alpha$ clustering on anisotropic flow. Our results demonstrate that the transport model successfully describes the measured collectivity in this intermediate-size system and that the magnitude and pattern of $v_{2}\{2\}$ and $v_{3}\{2\}$ serve as sensitive indicators of the underlying nuclear structure. Beyond the RHIC energy considered here, this unified framework provides a solid baseline for studying O+O collisions at other energies, enabling systematic investigations of geometry effects across different collision energies.

The paper is organized as follows. Section~\ref{sec:model} introduces the $^{16}$O nuclear configurations together with the improvements in the AMPT model. Section~\ref{sec:obs} outlines the observables and analysis methodology. Section~\ref{sec:Results} presents the main results and discussions. Finally, Section~\ref{sec:sum} summarizes the conclusions.

\section{\label{sec:model} Geometry configurations and improved AMPT model}

\begin{figure*}[htb]
  \begin{minipage}[t]{0.7\linewidth}
    \centering
    \includegraphics[width=0.9\textwidth]{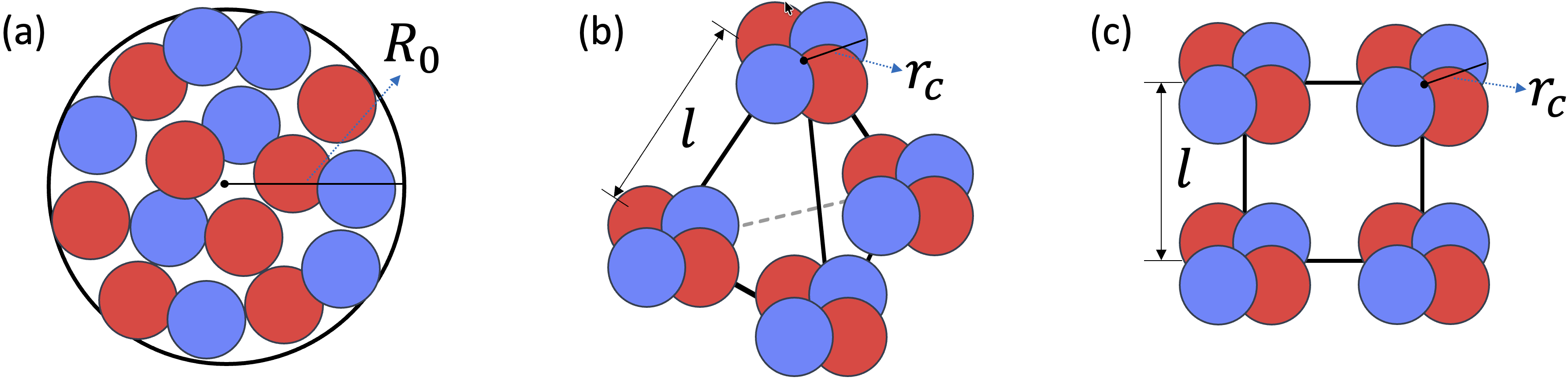}
     \end{minipage}
  \caption{Illustration of the geometrical configuration of $^{16}$O with the nucleon structures of (a) W-S distribution, (b) tetrahedron distribution, and (c) square distribution.}
\label{fig:cluster}
\end{figure*}

\begin{figure}[htb]
  \begin{minipage}[t]{0.95\linewidth}
    \centering
    \includegraphics[width=0.9\textwidth]{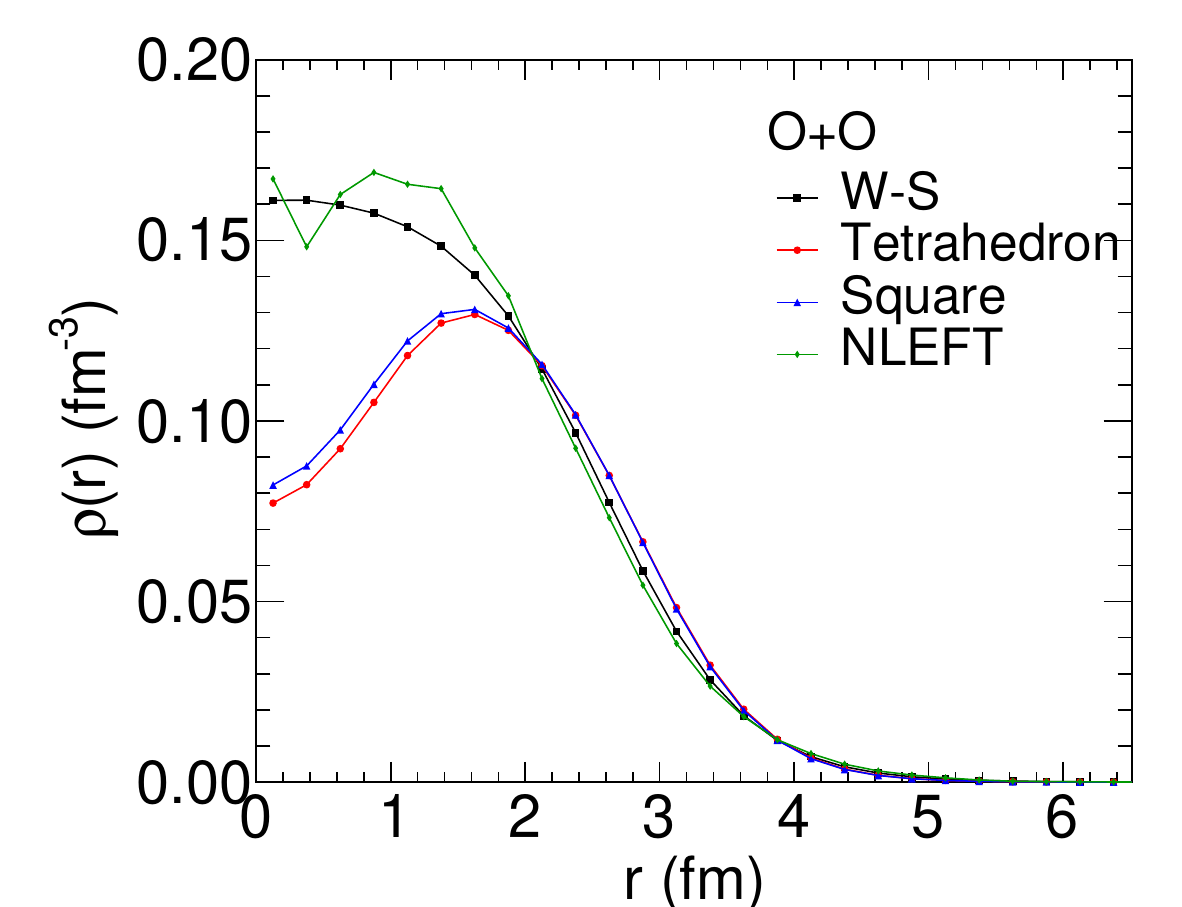}
  \end{minipage}
  \caption{The nucleon density distribution inside $^{16}$O in O+O collisions with the geometrical configurations of W-S, tetrahedron, square, and NLEFT distributions.}
\label{fig:rho}
\end{figure}

\begin{figure}[htb]
  \begin{minipage}[t]{0.95\linewidth}
    \centering
    \includegraphics[width=0.9\textwidth]{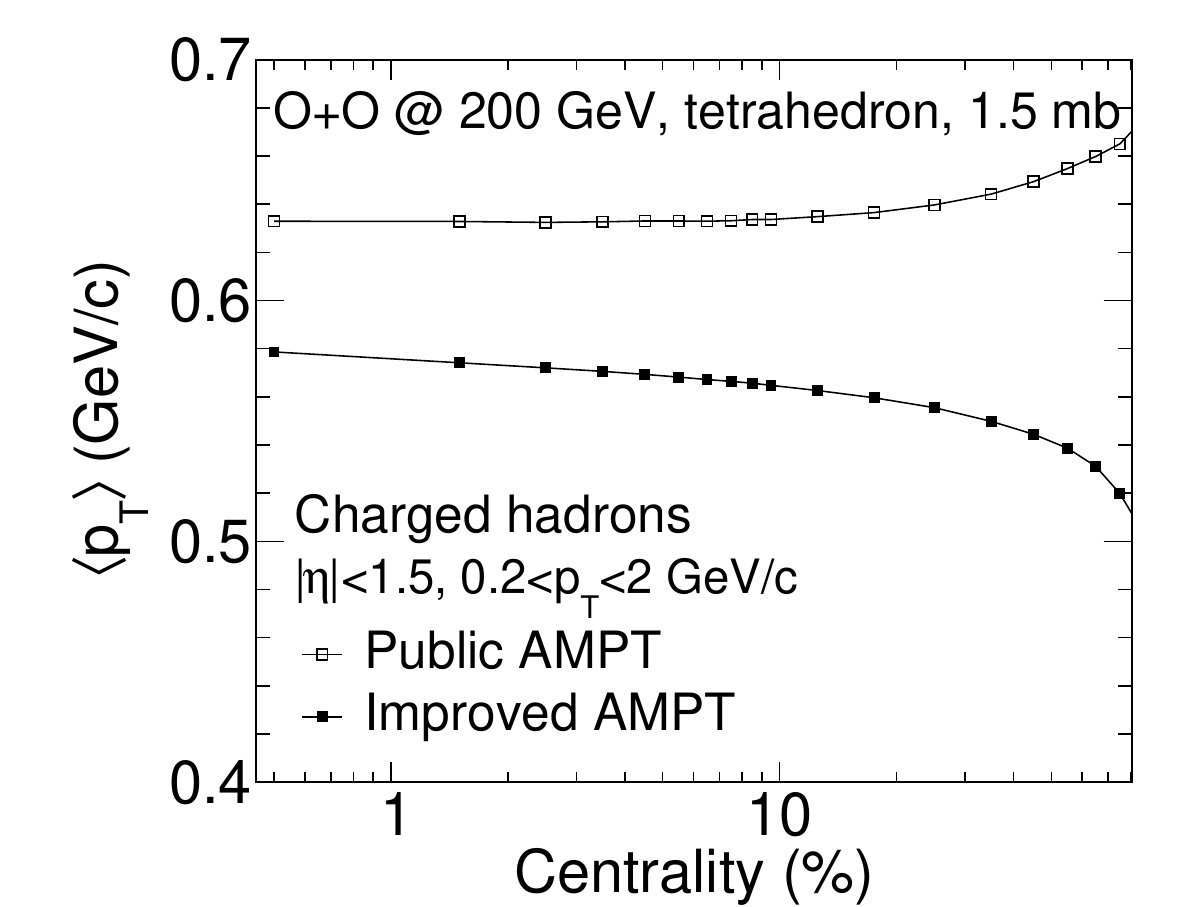}
  \end{minipage}
  \caption{The centrality dependence of $\left<p_{\rm T} \right>$ in O+O collisions at 200 GeV with the tetrahedron configuration of nuclear structure for two versions of the AMPT model with parton cross section 1.5 mb.}
\label{fig:mpt}
\end{figure}

\begin{figure}[htb]
  \begin{minipage}[t]{0.95\linewidth}
    \centering
    \includegraphics[width=0.9\textwidth]{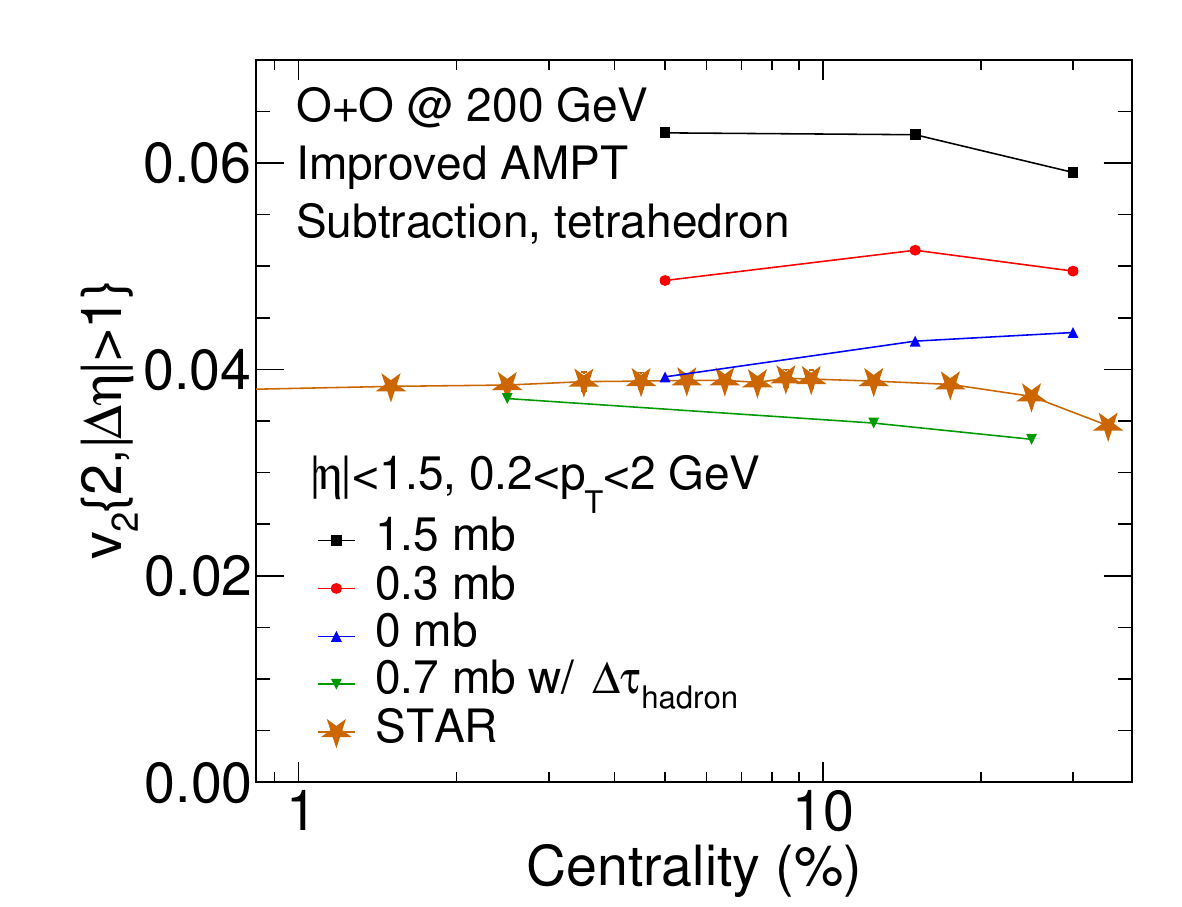}
  \end{minipage}
  \caption{The centrality dependence of $v_{2}\{2\}$ in O+O collisions at 200 GeV with different settings in the improved AMPT model for the tetrahedron configuration.}
\label{fig:v2diff}
\end{figure}

\begin{figure}[htb]
  \begin{minipage}[t]{0.95\linewidth}
    \centering
    \includegraphics[width=1.0\textwidth]{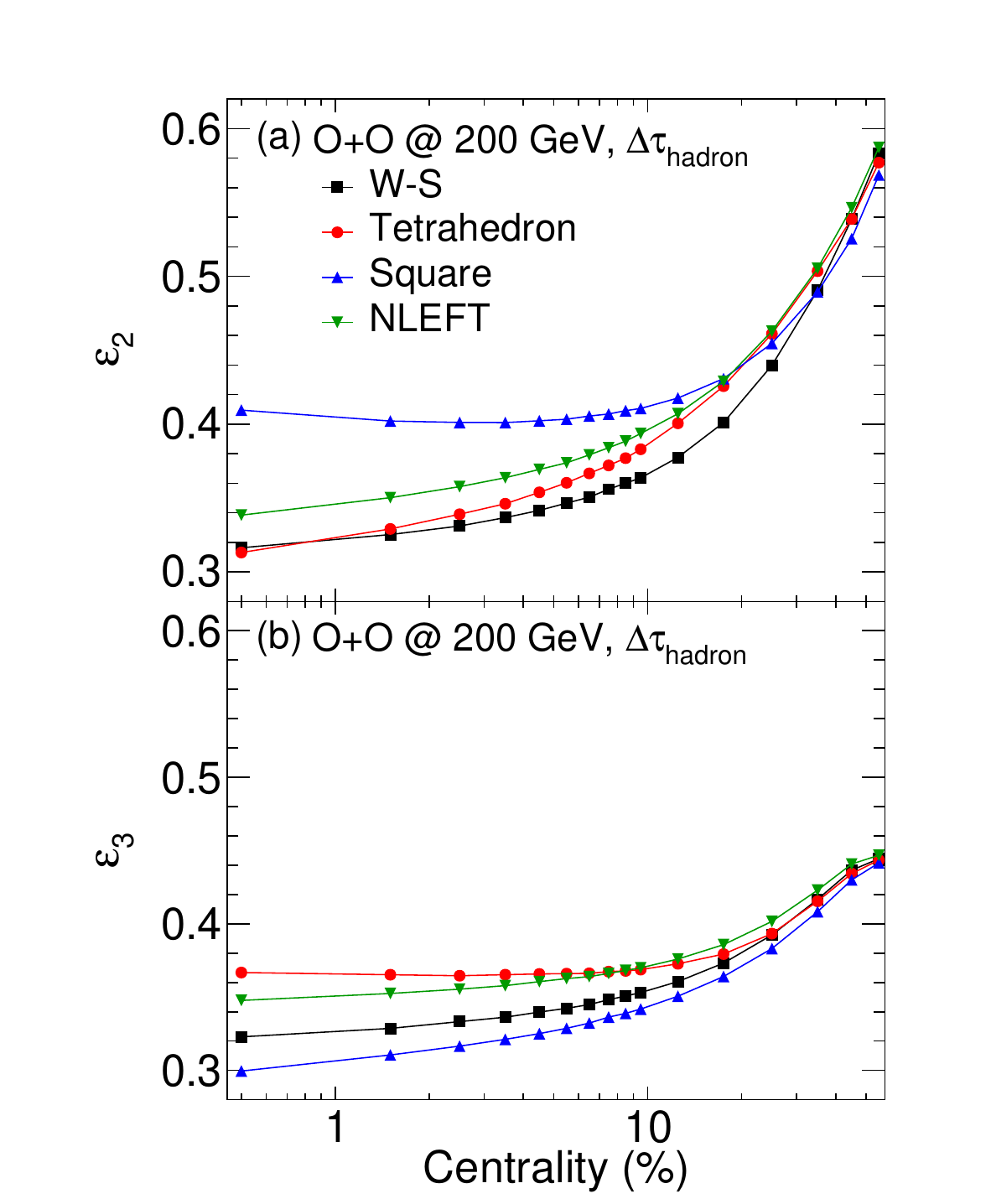}
  \end{minipage}
  \caption{The centrality dependence of (a) $\varepsilon_2$ and (b) $\varepsilon_3$ in O+O collisions at 200 GeV with different nuclear structure configurations.}
\label{fig:en}
\end{figure}

\begin{figure}
\begin{minipage}[t]{0.95\linewidth}
\subfigure{\includegraphics[width=0.9\textwidth]{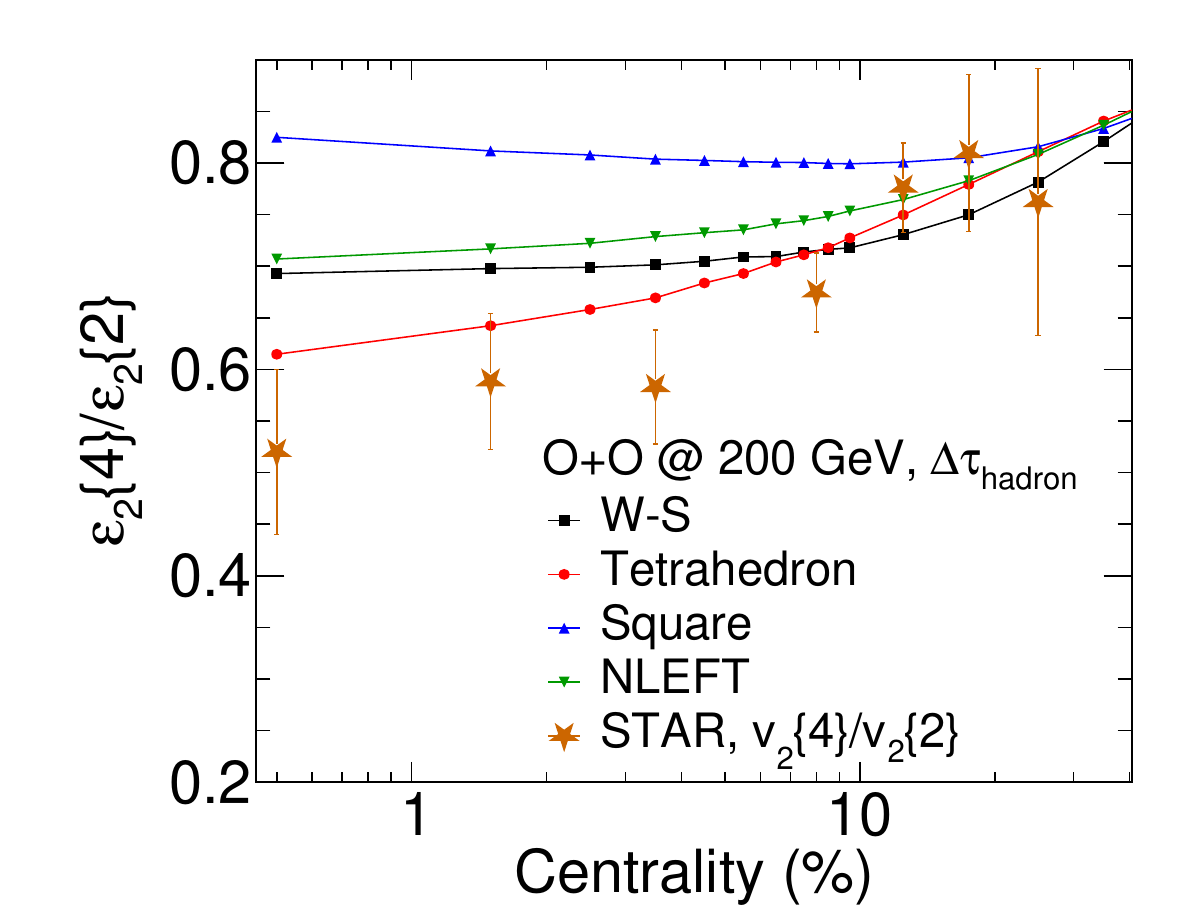}}
\end{minipage}
\caption{The centrality dependence of $\varepsilon_{2}\{4\}/\varepsilon_{2}\{2\}$ in O+O collisions at 200 GeV with different nuclear structure configurations, compared to STAR data.}
\label{fig:e24}
\end{figure}

\begin{figure*}[htb]
  \begin{minipage}[t]{0.33\linewidth}
    \centering
    \includegraphics[width=0.99\textwidth]{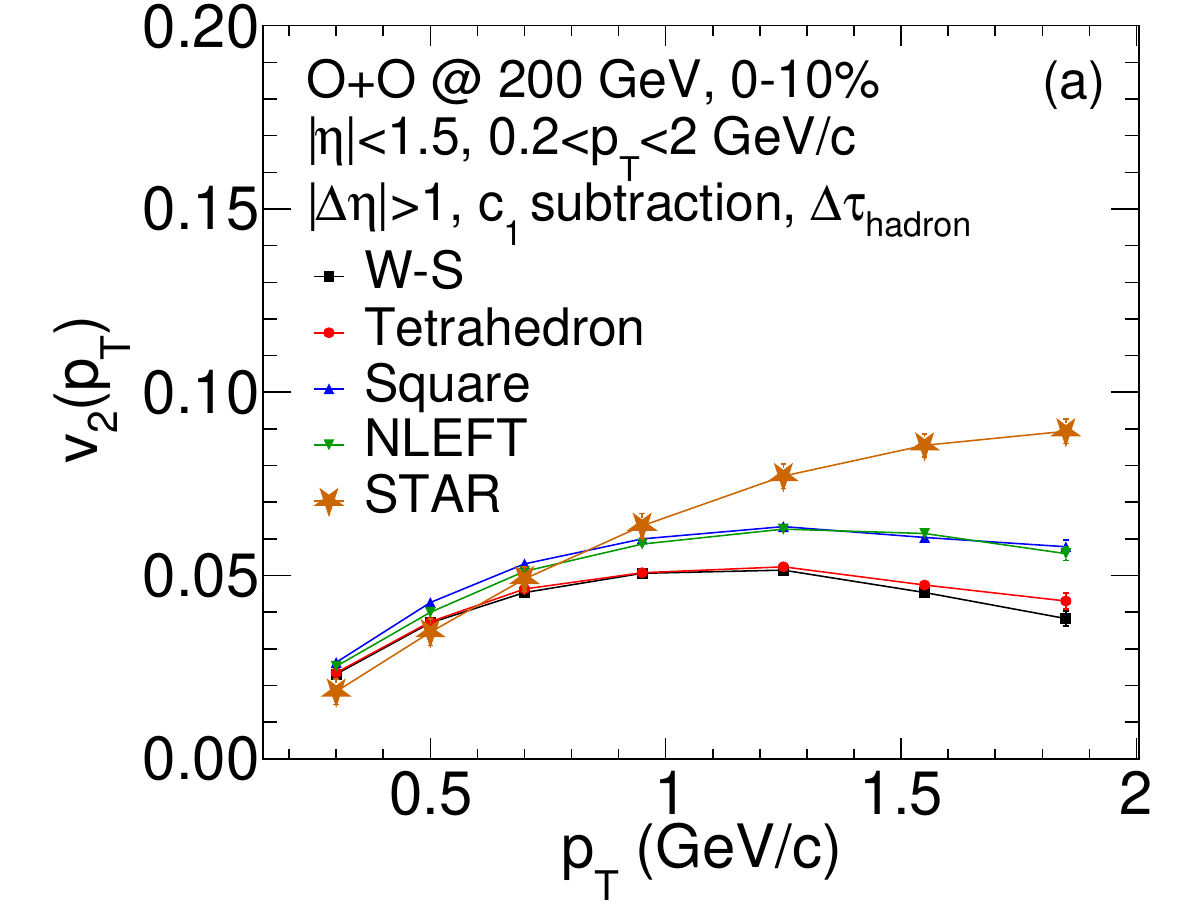}
  \end{minipage}%
     \begin{minipage}[t]{0.33\linewidth}
    \centering
    \includegraphics[width=0.99\textwidth]{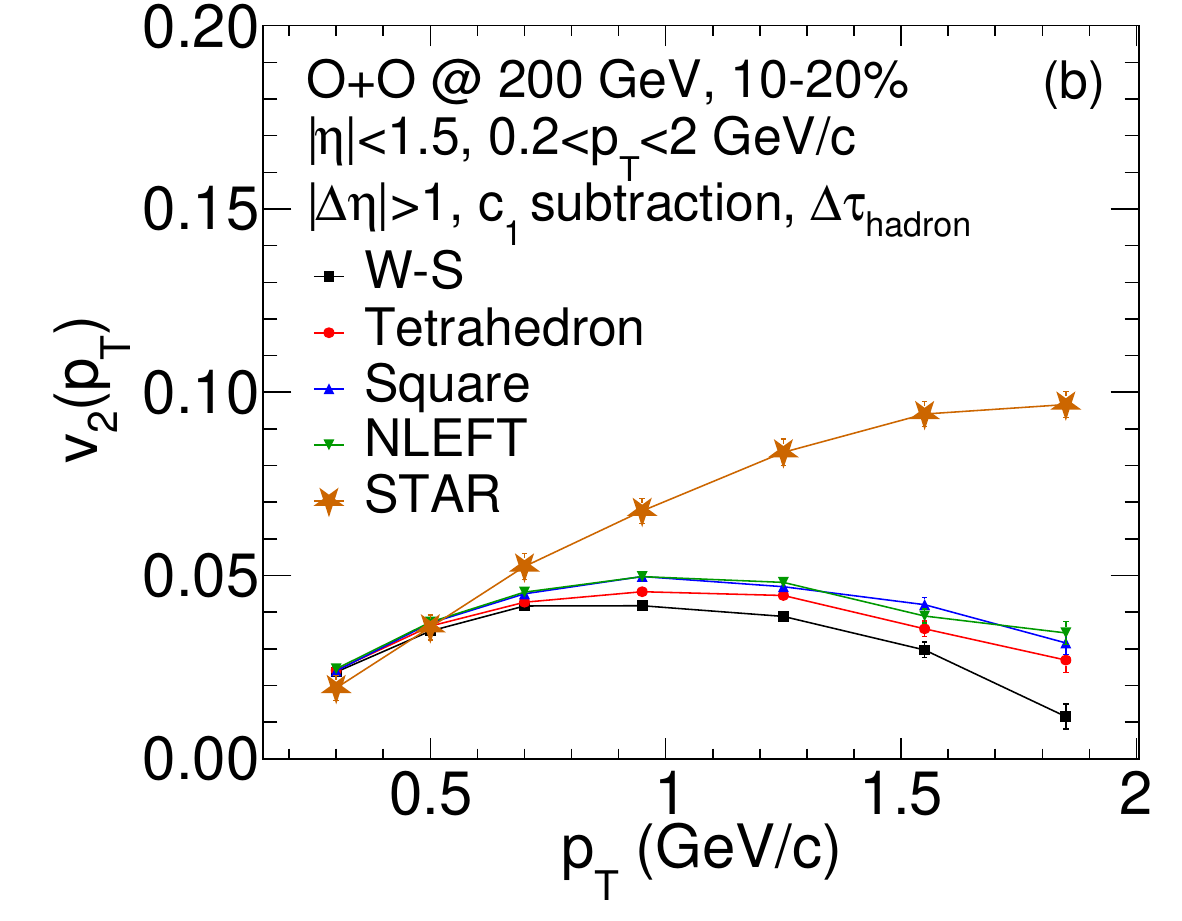}
     \end{minipage}%
     \begin{minipage}[t]{0.33\linewidth}
    \centering
    \includegraphics[width=0.99\textwidth]{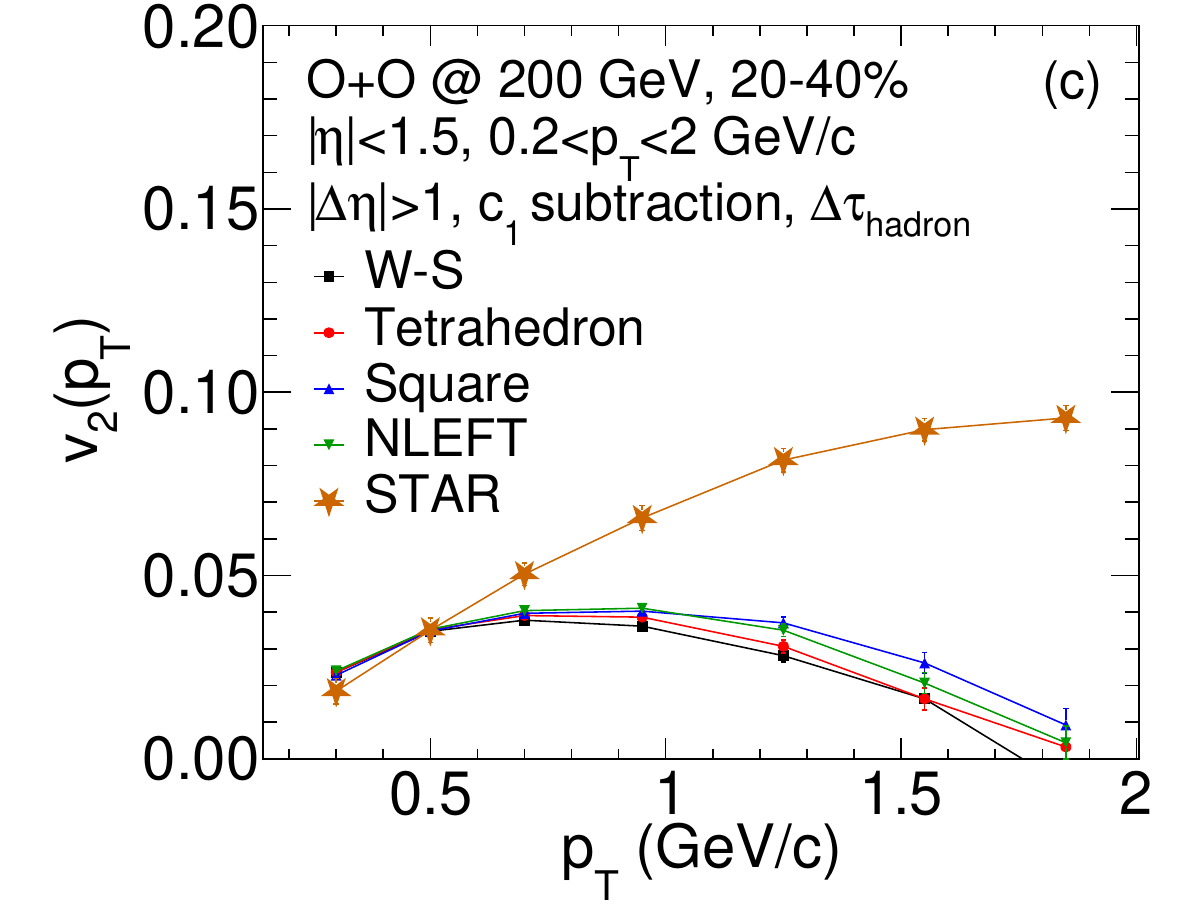} 
  \end{minipage}
  \caption{The $p_{\rm T}$ dependence of $v_{2}\{2\}$ for (a) 0-10\%, (b) 10-20\%, and (c) 20-40\% centrality bins in O+O collisions at 200 GeV with different nuclear structure configurations, compared to STAR data.}
\label{fig:v2pt}
\end{figure*}

\begin{figure*}[htb]
  \begin{minipage}[t]{0.33\linewidth}
    \centering
    \includegraphics[width=0.99\textwidth]{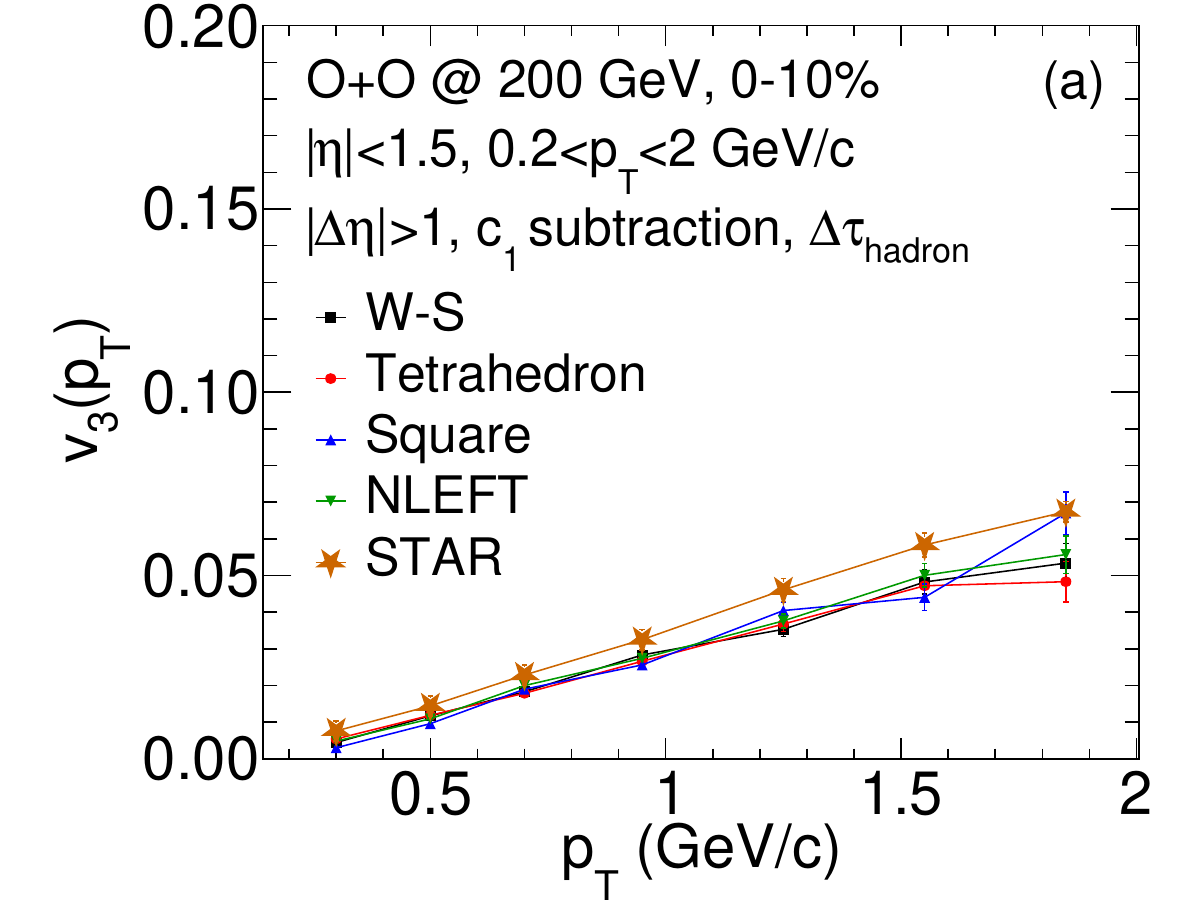}
  \end{minipage}%
     \begin{minipage}[t]{0.33\linewidth}
    \centering
    \includegraphics[width=0.99\textwidth]{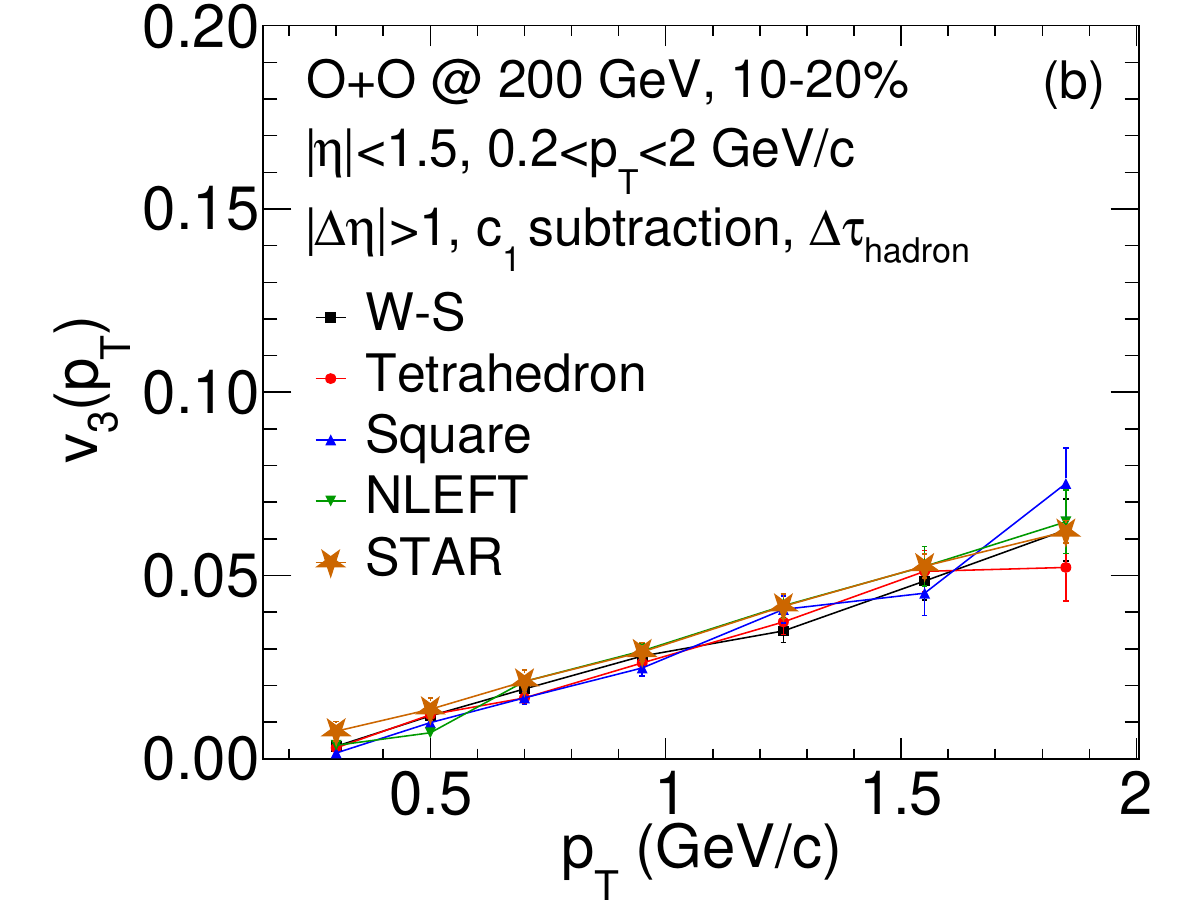}
     \end{minipage}%
     \begin{minipage}[t]{0.33\linewidth}
    \centering
    \includegraphics[width=0.99\textwidth]{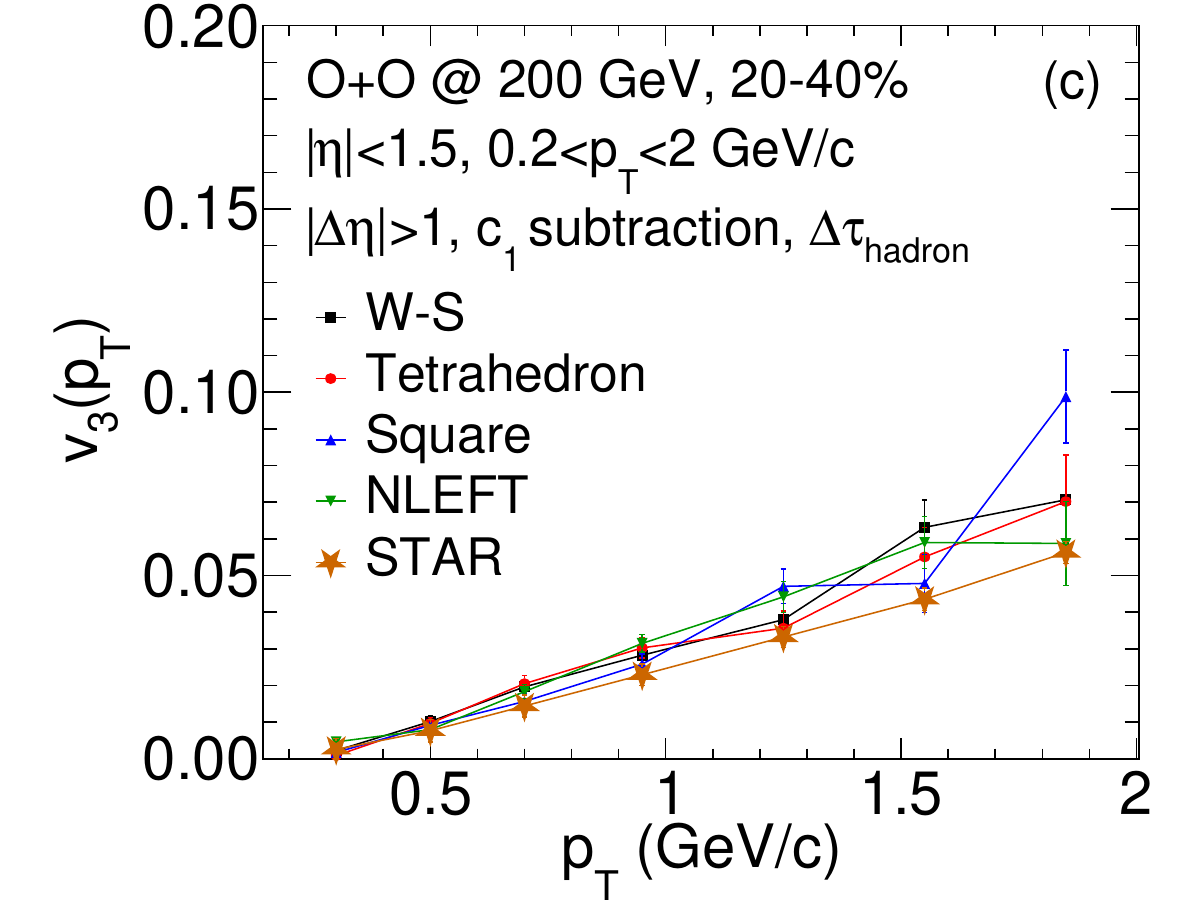} 
  \end{minipage}
    \caption{The $p_{\rm T}$ dependence of $v_{3}\{2\}$ for (a) 0-10\%, (b) 10-20\%, and (c) 20-40\% centrality bins in O+O collisions at 200 GeV with different nuclear structure configurations, compared to STAR data.}
\label{fig:v3pt}
\end{figure*}

\begin{figure}
\begin{minipage}[t]{0.95\linewidth}
\subfigure{\includegraphics[width=1\textwidth]{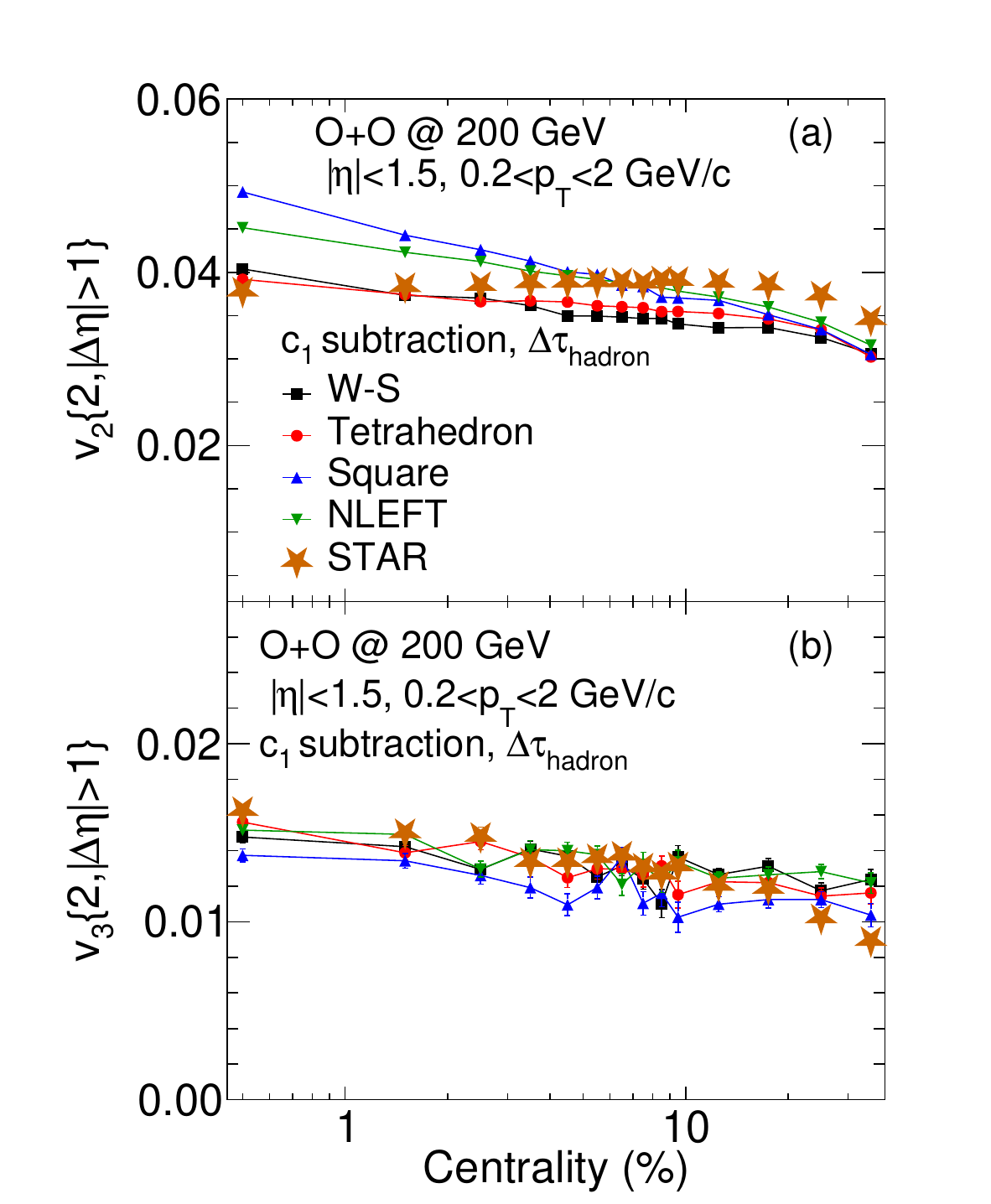}}
\end{minipage}
\caption{The centrality dependence of (a) $v_{2}\{2\}$ and (b) $v_{3}\{2\}$ in O+O collisions at 200 GeV with different nuclear structure configurations, compared to STAR data.}
\label{fig:vn}
\end{figure}

Applying the experimental root-mean-square ($rms$) nuclear charge radius $\left \langle r^{2} \right \rangle ^{1/2}=2.6991$ fm for $^{16}$O~\cite{Angeli:2013epw}, we chose suitable parameters for the four nuclear structures of $^{16}$O as follows:

(a) For the W-S distribution as shown in Fig.~\ref{fig:cluster}(a), the three-parameter Fermi (3pF) model~\cite{DeVries:1987atn} is used 
\begin{eqnarray}
\rho(r)=\rho_0(1+wr^2/R_0^2)/(1+{\rm exp}(\frac{r-R_0}{a})),
 \label{eq:rho}
\end{eqnarray}%
where $\rho_0$ is the nuclear saturation density, the radius of the nucleus $R_0=2.608$ fm, the surface diffusion parameter $a=0.513$ fm, and the weight parameter of the 3pF model $w=-0.051$.

(b) For the tetrahedron~\cite{Bauhoff:1984zza,Liu_2012,Girod:2013faa,Epelbaum:2013paa,Bijker:2014tka} cluster structure, the nucleon distributions are arranged as shown in Fig.~\ref{fig:cluster}(b). The distance between the centers of the clusters is $l$. Nucleons within each cluster are sampled from a Gaussian distribution:
\begin{eqnarray}
f_{i}({\textbf{\textit{r}}})\propto {\rm exp} (-\frac{3}{2}\frac{(\textbf{\textit{r}}-\textbf{\textit{c}}_{i})^2}{r_{c}^{2}}),
\label{eq:cfr}
\end{eqnarray}%
where $\textbf{\textit{r}}$ is the coordinate of the nucleon, $\textbf{\textit{c}}_{i}$ is the center position of cluster $i$, and $r_{c}$ is the $rms$ radius of a cluster. As shown in Fig.~\ref{fig:cluster}(b), $l=3.5$ fm and $r_c=1.23$ fm.

(c) For the square cluster structure, the nucleons are arranged as shown in Fig.~\ref{fig:cluster}(c). Nucleons within each cluster are sampled as Eq.~(\ref{eq:cfr}), and $l=3.0$ fm, $r_c=1.23$ fm. 

(d) The NLEFT case combines effective field theory with lattice Monte Carlo methods to generate ground-state nuclear configurations that include many-body correlations at all orders. In this study, we employ a minimal pion-less EFT Hamiltonian on a periodic lattice~\cite{Summerfield:2021oex,Giacalone:2024luz}, which successfully reproduces binding energies and charge radii of the relevant isotopes.

Figure~\ref{fig:rho} presents the nucleon density distributions for the above-mentioned four configurations. It can be observed that the nucleon density distributions for the W-S and NLEFT configurations are significantly higher than those of the tetrahedron and square configurations. This characteristic can be inferred from Fig.~\ref{fig:cluster}. 
The NLEFT curve is generated from sampled nucleon coordinates that are re-centered on a nucleus-by-nucleus basis. This re-centering introduces a slight bias in the $\rho(r)$ profile, as discussed in Ref.~\cite{Jia:2022qgl}.
For the W-S configuration, the nucleons are concentrated more toward the center of the nucleus, while for the tetrahedron and square configurations, there are fewer nucleons.

Next, the above four configurations of $^{16}$O are introduced into the initial conditions of the string-melting version of AMPT (AMPT-SM) model~\cite{Lin:2004en} to describe the evolution of QGP and hadrons in high-energy nuclear collisions~\cite{Ma:2016fve,Chen:2022xpm,Zhang:2021kxj,Zhao:2022grq}.  
The AMPT model incorporates initial conditions derived from the HIJING model, parton interactions as described in the Zhang’s Parton Cascade (ZPC) model, quark coalescence for hadronization, and hadron interactions using A Relativistic Transport (ART) model. In the AMPT-SM model, the initial partons are generated by string melting after a formation time of $t_{\rm p}=E_{\rm h}/m_{\rm T,h}^{2}$, where $E_{\rm h}$ and $m_{\rm T,h}$ represent the energy and transverse mass of the parent hadron. The parent hadrons are produced at the transverse positions of either the projectile or the target participant nucleons. The positions of the partons from the melting of parent hadrons are determined based on the positions of their parent hadrons using straight-line trajectories.  Similarly, hadrons produced by coalescence are assigned a proper formation time of $\tau_{\rm hadron}$=0.7 fm/$c$ in their rest frame prior to participating in hadronic rescatterings.
Unlike publicly available AMPT-SM models, including many previous AMPT studies on nuclear structure~\cite{Li:2020vrg,Behera:2023nwj,Jia:2022qgl,Zhang:2021kxj}, here we use the latest improved AMPT-SM model~\cite{Zhang:2019utb,Zheng:2019alz,Zhang:2021vvp,Lin:2021mdn,Zhang:2022fum,Zhang:2023xkw} for the first time.
This improved AMPT-SM model has implemented a new quark coalescence~\cite{He:2017tla}, modern parton distribution functions for the free proton, impact parameter-dependent nuclear shadowing, and improved heavy flavor production. The improved model can provide more reasonable descriptions of the particle yield ratios, $p_{\rm T}$ spectra, and anisotropic flows in $\rm A+A$ collisions at RHIC and LHC energies~\cite{Zhang:2019utb,Zheng:2019alz,Zhang:2021vvp,Lin:2021mdn,Zhang:2022fum,Zhang:2023xkw}. 
For example, as shown in Fig.~\ref{fig:mpt}, the public AMPT-SM model with a parton cross section of 1.5 mb yields an increase in $\left<p_{\rm T}\right>$ from central to peripheral collisions in O+O collisions at 200 GeV, which is contrary to the expected trend. 
In contrast, the improved AMPT-SM model exhibits the expected centrality dependence, owing to the recent incorporation of local nuclear scaling in the initial conditions~\cite{Zhang:2021vvp}. 
In this work, we employ the improved AMPT-SM model for the first time to investigate the nuclear structure in high-energy nuclear collisions. 

The impact of partonic cross section on $v_{2}\{2\}$ has also been investigated, presented in Fig.~\ref{fig:v2diff}. Our findings indicate that obtaining results similar to those of STAR data is challenging. Even when the parton cross section is set to zero, the model still produces a too large $v_{2}\{2\}$. This indicates that the hadronic interactions in O+O collisions play a dominant role in generating $v_{2}\{2\}$ within the AMPT-SM framework. Unlike hydrodynamic models, where hadronization is governed by local energy density, the hadronization in the AMPT-SM model occurs when no further scatterings between partons happen. In the Appendix~\ref{sec:tauhampt}, our analysis reveals that the AMPT-SM model produces hadrons too early in O+O collisions in that the peak energy density of the hadron phase often well exceeds the expected energy density for the QCD phase transition. To solve this issue, we delay the hadron formation time in our study so that the peak energy density of the hadronic matter in the center cell of O+O collisions at different centralities is around $\rm 0.3~GeV/fm^3$. With this setup, the improved AMPT-SM model which has a parton scattering cross section of 0.7~mb reasonably reproduces the $v_{2}\{2\}$ measurement from STAR, as indicated by the green line of the AMPT model in Fig.~\ref{fig:v2diff}. Note that we have not included heavy flavor hadrons in this improvement, and all the presented $v_n$ results have excluded the contributions from heavy flavor hadrons that carry charm and bottom hadrons.

\section{\label{sec:obs} Observables and methodology}

In this paper, the initial eccentricity $\varepsilon_{2}$ and triangularity $\varepsilon_{3}$ are calculated from the spatial distribution of all initial partons. The initial parton refers to a parton at its formation time, which is determined by the energy and transverse mass of its parent hadron. A parton is assumed to free-stream from its production point during its formation time. The definition of $\varepsilon_{n}$ is as follows~\cite{Ma:2016hkg}:
\begin{eqnarray}
\varepsilon_{n} = 
\frac{\sqrt{\langle r^{n}\cos(n\varphi)\rangle^{2} + \langle r^{n}\sin(n\varphi)\rangle^{2}}}
{\langle r^{n}\rangle},
\label{eq:e23}
\end{eqnarray}%
here $r$ and $\varphi$ represent the radial position and azimuthal angle of each initial parton in the transverse plane, while 
${\langle ...\rangle}$ denotes a particle-weighted average. The two- and four-particle cumulants of $\varepsilon_{2}$ are defined as
\begin{eqnarray}
\varepsilon _{2}^{2}\{2\}&=&\left \langle \varepsilon_{2}^{2}  \right \rangle = \langle \varepsilon_{2} \rangle^{2} + \sigma_{\varepsilon_2}^{2},
\label{eq:e22}
\\
\varepsilon _{2}^{2}\{4\}&=&(-\left \langle \varepsilon_{2}^{4}  \right \rangle +2\left \langle \varepsilon_{2}^{2}  \right \rangle^{2} )^{1/2} \approx \langle \varepsilon_{2} \rangle^{2} - \sigma_{\varepsilon_2}^{2},
 \label{eq:e24}
\end{eqnarray}%
where $\sigma_{\varepsilon_2}^{2}$ represents the variance of the event-by-event distribution of $\varepsilon_2$.

The study of anisotropic flow in O+O collisions is performed using the two-particle correlation method~\cite{Huang:2023viw}.
By using the Fourier expansion of the two-particle correlation,
\begin{eqnarray}
\frac{dN^{\rm pairs}}{d\Delta\phi} \propto 1+2\sum_{n=1 }^{\infty} v_{n}{\rm cos}(n\Delta\phi),
\label{eq:dN}
\end{eqnarray}%
the anisotropic flow can be extracted. Here the $\Delta \phi$ is the relative azimuthal angle between the trigger particle $\phi^{\rm trig}$ and the associated particle $\phi^{\rm assoc}$, i.e. $\Delta \phi =\phi^{\rm trig}-\phi^{\rm assoc}$.  $v_{n}$ is the coefficient of the $n$th-order flow. To obtain $v_{n}(p_{\rm T}^{\rm trig})$, the Fourier fitting can be employed on the $Y(\Delta\phi)$ distribution:
\begin{eqnarray}
Y(\Delta\phi,p_{\rm T}^{\rm trig})=c_{0}(1+2\sum_{n=1}^{n=4} c_{n}{\rm cos}(n\Delta\phi)),
\label{eq:dY}
\end{eqnarray}%
$c_{n}$ is the product of $v_n$ for the trigger and associated particles, $c_{n}=v_{n}^{\rm trig} \times v_{n}^{\rm assoc}$. Then, the non-flow contamination can be subtracted following
\begin{eqnarray}
c_{n}^{\rm sub}=c_{n}-c_{n}^{\rm non-flow}=c_{n}-c_{n}^{\rm peri}\times f,
\label{eq:vn-sub}
\end{eqnarray}%
where $f=c_{1}/c_{1}^{\rm peri}$ ($c_{1}$ subtraction). In this work, the anisotropic flow is extracted using Fourier fitting and non-flow subtraction with $60-80\%$ peripheral collisions, as in the STAR measurement~\cite{Huang:2023viw}. The $c_{1}$ subtraction is found to be equivalent to the alternative normalization $f=c_{0}^{\rm peri}/c_{0}$ ($c_{0}$ subtraction). The STAR data show that both methods give consistent $f$ values within uncertainties. 
However, when applied to AMPT model calculations, the two normalization schemes lead to slightly different results. By default, we present the results based on the $c_{1}$ subtraction for consistency and clarity in this paper. The relevant results using the $c_{0}$ subtraction method) are provided in the Appendix~\ref{sec:diffnonflow} for completeness and comparison.

\section{\label{sec:Results} Results and discussions}

Figure~\ref{fig:en} (a) and (b) present the AMPT-SM results on the centrality dependence of $\varepsilon_{2}$ and $\varepsilon_{3}$ in O+O collisions at 200 GeV for the W-S, tetrahedron, square, and NLEFT configurations. Both $\varepsilon_{2}$ and $\varepsilon_{3}$ increase from central to peripheral collisions. In addition, $\varepsilon_{2}$ is the largest for the square configuration and the smallest for the W-S configuration in central and mid-central collisions. 
The enhanced $\varepsilon_{2}$ in the square configuration stems from its strongly oblate shape, which is equivalent to a large negative $\beta_{2}$ quadrupole deformation. This shape intrinsically boosts the ellipticity compared to more spherical arrangements.
Meanwhile, $\varepsilon_{3}$ is the largest for the tetrahedron configuration and the smallest for the square configuration at central and mid-central collisions. 
In peripheral collisions, the small overlapping region leads to fewer participating nucleons. In this regime, the influence of intrinsic nuclear geometries, such as tetrahedron or NLEFT structures, on initial spatial anisotropy is negligible.
The following joint investigation of $v_{2}\{2\}$ and $v_{3}\{2\}$ has the potential to provide independent and complementary constraints on the underlying nuclear structure.
It is worth noting that in this study, the centrality in this work is defined according to the multiplicity distribution of charged particles within $|\eta|<1.5$ and $0.2<p_{\rm T}<2$ GeV/$c$.

Considering that the final state anisotropic flow is driven by the initial state geometry, the study of anisotropic flow and its event-by-event fluctuations provides insight into the initial geometry and its fluctuations~\cite{Jia:2022qgl,Zhang:2021kxj,STAR:2024wgy,Giacalone:2024luz,Giacalone:2023cet}. Recent works have shown that the sensitivity of flow observables to the initial geometry can reveal additional information on the underlying nuclear structure in ultra-relativistic collisions~\cite{Jia:2021qyu, Lu:2023fqd, Nielsen:2023znu}. In this study, we investigate the structure of $^{16}$O by examining the initial-state eccentricity fluctuations, quantified through the ratio $\varepsilon_{2}\{4\}/\varepsilon_{2}\{2\}$.
As indicated in Eqs.~(\ref{eq:e22}) and (\ref{eq:e24}), a larger $\sigma_{\varepsilon_2}$ leads to a larger deviation of the $\varepsilon_{2}\{4\}/\varepsilon_{2}\{2\}$ ratio from unity. The centrality dependence of $\varepsilon_{2}\{4\}/\varepsilon_{2}\{2\}$ ratio is shown in Fig.~\ref{fig:e24}. 
We find that for the square configuration, the $\varepsilon_{2}\{4\}/\varepsilon_{2}\{2\}$ ratio below 10\% centrality decreases as centrality increases. Conversely, the $\varepsilon_{2}\{4\}/\varepsilon_{2}\{2\}$ ratios for the tetrahedron, NLEFT and W-S configurations exhibit modest increases with centrality, while the results from the W-S configuration show a weak centrality dependence for the central collision 0-10\% centrality. Furthermore, the $\varepsilon_{2}\{4\}/\varepsilon_{2}\{2\}$ ratio for the square configuration is notably higher than those for the other configurations. This behavior can be understood as a consequence of its strong oblate deformation, which enhances higher-order cumulants more efficiently than the two-particle cumulant. This difference among the four configurations is most pronounced in central collisions and diminishes toward more peripheral collisions. 
In addition, when comparing the $\varepsilon_{2}\{4\}/\varepsilon_{2}\{2\}$ ratios with the STAR measurements on $v_{2}\{4\}/v_{2}\{2\}$, we found that the results from the tetrahedron give better descriptions of the STAR measurements than the calculations using NLEFT, W-S and square configurations.

Measurements of $v_{n}(p_{\rm T})$ for charged and identified particles offer a window into the initial geometry and its fluctuations in the QGP. Although the integrated $v_{n}$ is sensitive to the overall anisotropy and nuclear structure, the $p_{\rm T}$-dependent $v_{n}(p_{\rm T})$ captures finer details of the initial-state spatial configurations.
Figures~\ref{fig:v2pt} (a)-(c) show the $p_{\rm T}$ dependence of $v_{2}\{2\}$ for (a) 0-10\%, (b) 10-20\%, and (c) 20-40\% centrality bins, respectively, in O+O collisions at 200 GeV from the AMPT-SM model. The calculations from four configurations, together with the comparisons to the STAR measurements~\cite{Huang:2023viw}, are presented. We can see that $v_{2}(p_{\rm T})$ calculations first increase and then decrease with $p_{\rm T}$ for all three centrality bins and for all configurations of the AMPT-SM model. The calculations can describe STAR measurements for $p_{\rm T}<0.6$ GeV/$c$ but underestimate the $v_{2}(p_{\rm T})$ measurements for $p_{\rm T} > 1$ GeV/$c$. We also observe clear differences among configurations in 0-10\% central collisions. The square configuration shows the largest $v_{2}(p_{\rm T})$, whereas the W-S configuration shows the smallest $v_{2}(p_{\rm T})$. This can be understood because the square configuration results in a maximum of $\varepsilon_{2}$ in the central collision, while the W-S configuration results in a minimum of $\varepsilon_{2}$, as shown in Fig.~\ref{fig:en} (a).

Similarly, the $p_{\rm T}$ dependence of $v_{3}\{2\}$ in O+O collisions at 200 GeV from the AMPT-SM model with different configurations are shown in Fig.~\ref{fig:v3pt}. The results are presented in (a) 0-10\%, (b) 10-20\% and (c) 20-40\%, respectively. It can be seen that $v_{3}\{2\}$ increases linearly as $p_{\rm T}$ increases. The $p_{\rm T}$-dependent $v_{3}\{2\}$ results from the AMPT-SM model with different configurations show compatible results within uncertainties. 
All these calculations are compatible with the STAR measurements. 
The near-independence of $v_3(p_{\rm T})$ on different $^{16}$O configurations stems from the facts that $v_{3}\{2\}$ is primarily driven by fluctuation-induced triangularity $\varepsilon_3$, its response to the initial geometry is relatively weak, and residual non-flow and finite-multiplicity effects further reduce the sensitivity of $v_{3}\{2\}$. Thus, while $v_{2}\{2\}$ remains sensitive to configuration-dependent ellipticity, the successful description of $v_{3}\{2\}$ by AMPT-SM demonstrates that the model captures the essential fluctuation-driven dynamics. This provides a reliable baseline for probing nuclear structure and clustering effects.

Finally, we analyze the $p_{\rm T}$-integrated $v_{n}$ versus centrality to quantify the overall geometric effects and enable direct comparison with STAR measurements.
Figures~\ref{fig:vn}(a) and (b) present the centrality dependence of $v_{2}\{2\}$ and $v_{3}\{2\}$ in O+O collisions at 200 GeV from the AMPT-SM model. Results from the four configurations are shown, together with the comparisons to the STAR measurements~\cite{Huang:2023viw}. In Fig.~\ref{fig:vn} (a), the $v_{2}\{2\}$ results from the AMPT-SM model are close to the $v_{2}\{2\}$ measurement from STAR, despite the fact that the AMPT-SM results exhibit a stronger centrality dependence, especially for the square configuration.
When comparing the AMPT-SM calculations from the four configurations, we find that the difference in $v_{2}\{2\}$ is most pronounced in central collisions, where the square configuration shows the largest $v_{2}\{2\}$ and the W-S configuration shows the smallest $v_{2}\{2\}$, which is consistent with the $p_{\rm T}$-dependent $v_{2}\{2\}$ results in Fig.~\ref{fig:v2pt}. In Fig.~\ref{fig:vn} (b), the $v_{3}\{2\}$ results from the AMPT-SM model well reproduce the STAR data. The $v_{3}\{2\}$ result for the NLEFT configuration is the largest among all configurations.
These results collectively demonstrate that both the formation of anisotropic flow and its fine structures can be significantly modulated by the initial nuclear configuration, offering a novel approach to probing nuclear cluster structures in high-energy collisions.

\section{\label{sec:sum} Conclusion}
In this work, we employed the improved AMPT-SM transport model to study the impact of different $^{16}$O nuclear configurations on anisotropic flows in O+O collisions at $\sqrt{s_{_{\rm NN}}}=200$~GeV. Using a single model setup, we considered four candidate geometries: Woods–Saxon, tetrahedron, square, and NLEFT, enabling a direct assessment of how nuclear structure and possible $\alpha$ clustering influence flow observables. 
To characterize the initial geometry, we computed the eccentricity cumulant ratio $\varepsilon_{2}\{4\}/\varepsilon_{2}\{2\}$ from the initial density distributions. This ratio predominantly reflects geometry-driven fluctuations of the initial state and is sensitive to different $^{16}\mathrm{O}$ configurations.
Regarding final-state observables, the $p_{\rm T}$-dependent $v_{2}\{2\}$ is well described at low $p_{\rm T}$, with underestimation at higher $p_{\rm T}$, while $v_{3}(p_{\rm T})$ agrees with STAR data across the full $p_{\rm T}$ range. Integrated $v_{2}\{2\}$ and $v_{3}\{2\}$ values also match experimental magnitudes, demonstrating that the transport framework captures the essential collectivity in this intermediate-size system and that anisotropic flow observables effectively probe nuclear geometry and possible $\alpha$ clustering. 
Overall, our study demonstrates that nuclear geometry has a significant impact on anisotropic flow, and the improved AMPT model can effectively describe STAR measurements in O+O collisions. These results provide a solid foundation for extending similar studies to other collision energies and for exploring exotic nuclear structures through relativistic heavy-ion collisions. Currently, the AMPT results only include  statistical uncertainties. In future work, we plan to incorporate systematic uncertainties using a Bayesian approach to further refine and quantify the model predictions.

The present work, utilizing the latest developments of the AMPT-SM model, indicates that anisotropic flow observables are sensitive probes of cluster configurations in $^{16}$O nuclei. The agreement with STAR data is achieved through a phenomenological adjustment to the hadron formation time, enabling a quantitative description of the collectivity in this intermediate-size system. 
Building on these results, future studies can leverage existing O+O flow measurements from RHIC and the LHC at multiple energies to perform systematic, energy-dependent constraints on nuclear structure. Such efforts can be extended to include additional differential observables and higher-order cumulants within a unified transport-model framework. Moreover, future measurements in the fixed-target LHC program at LHCb using $^{16}$O isotopes~\cite{LHCb:2025abc} will provide valuable tests of our model predictions and further insights into geometry-driven collectivity.

\begin{acknowledgments}

We thank Prof. Bo Zhou and Dr. Chun-Jian Zhang for helpful discussions about cluster structure in $\rm ^{16}O$ and anisotropic flow in O+O collisions. In addition, we gratefully acknowledge Prof. Dean Lee and Prof. Christopher Plumberg for sharing the nucleon density distribution of oxygen. Finally, we acknowledge Dr. Chen Zhong for maintaining the high-quality performance of the Fudan supercomputing platform for nuclear physics. This work is supported by the National Natural Science Foundation of China under Grant No. 12105054 (X.Z.), the European Union (ERC, InitialConditions, 101077147) and Independent Research Fund Denmark grant No. 2064-00052 (Y.Z.), the National Natural Science Foundation of China under Grants No. 12325507, No. 12547102, and No. 12147101, and the National Key Research and Development Program of China under Grants No. 2022YFA1604900 (G.M.), the National Science Foundation under Grant No. 2310021 (Z.W.L.), the National Natural Science Foundation of China under Grant No.12405159 (C.Z.).

\end{acknowledgments}

\bibliography{refs-1}

\begin{appendices}
\titleformat{\section}
  {\normalfont\small\bfseries}
  {\appendixname~\thesection.}
  {0.5em} {}
\renewcommand{\thesection}{\Alph{section}}
\setcounter{section}{0} 
\section{\label{sec:tauhampt} Delay of hadron cascade in the AMPT-SM model}%

\begin{figure}[h]
\begin{minipage}[t]{0.95\linewidth}
\subfigure{\includegraphics[width=0.9\textwidth]{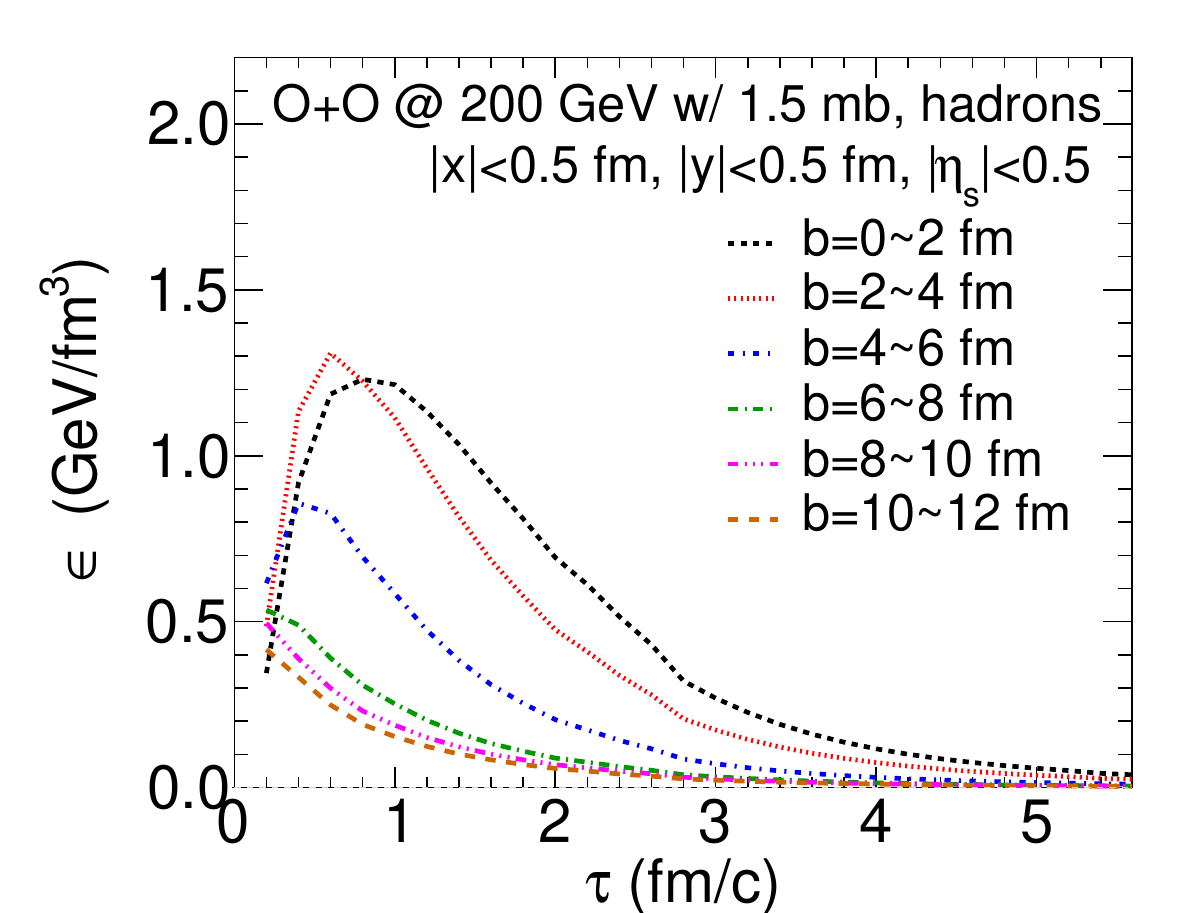}}
\end{minipage}
\caption{The evolution of energy density of hadrons in the center cell of the  AMPT model with parton cross section 1.5 mb before adding the extra hadron formation time.}
\label{fig:en0}
\end{figure}

\begin{figure}[h]
\begin{minipage}[t]{0.95\linewidth}
\subfigure{\includegraphics[width=0.9\textwidth]{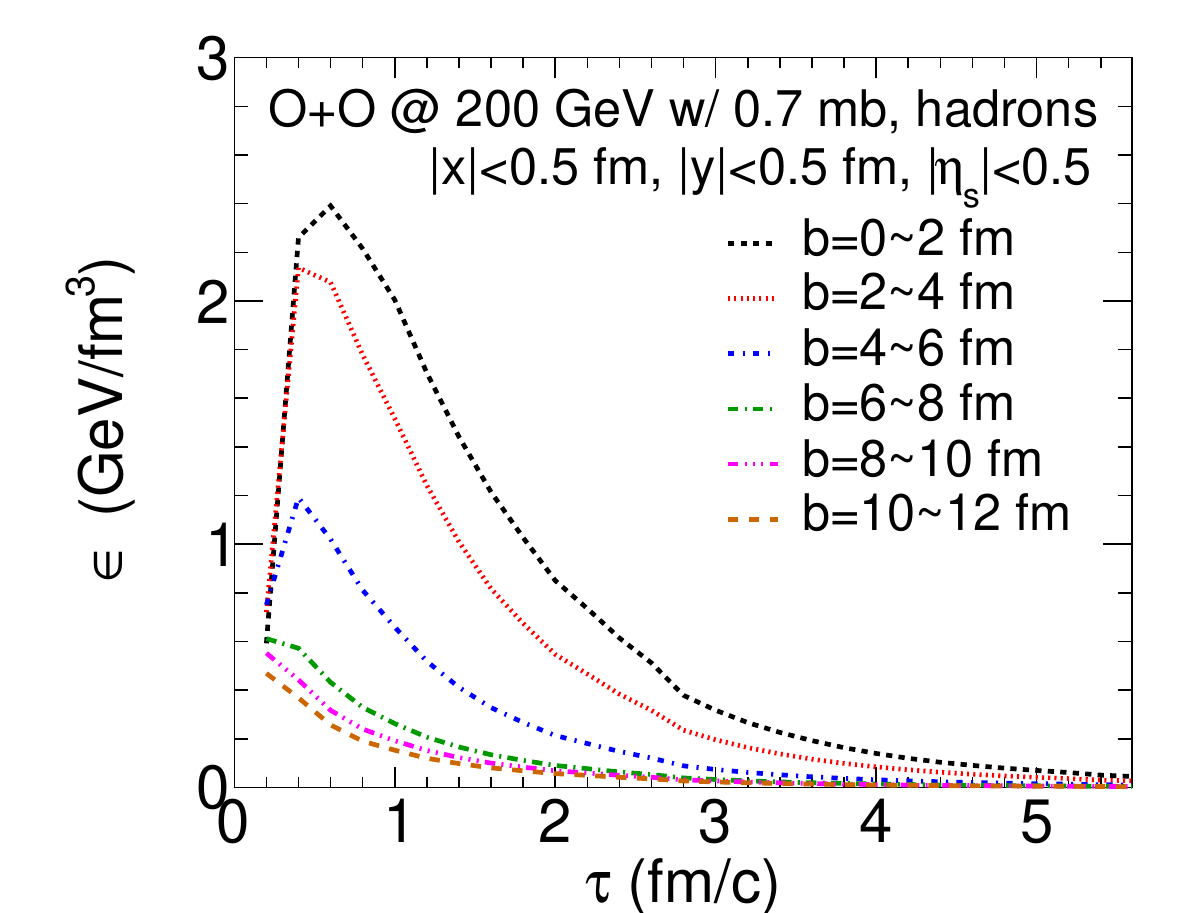}}
\end{minipage}
\caption{The evolution of energy density of hadrons in the center cell of the AMPT model with parton cross section 0.7 mb before adding the extra hadron formation time.}
\label{fig:en1}
\end{figure}

\begin{figure}[h]
\begin{minipage}[t]{0.95\linewidth}
\subfigure{\includegraphics[width=0.9\textwidth]{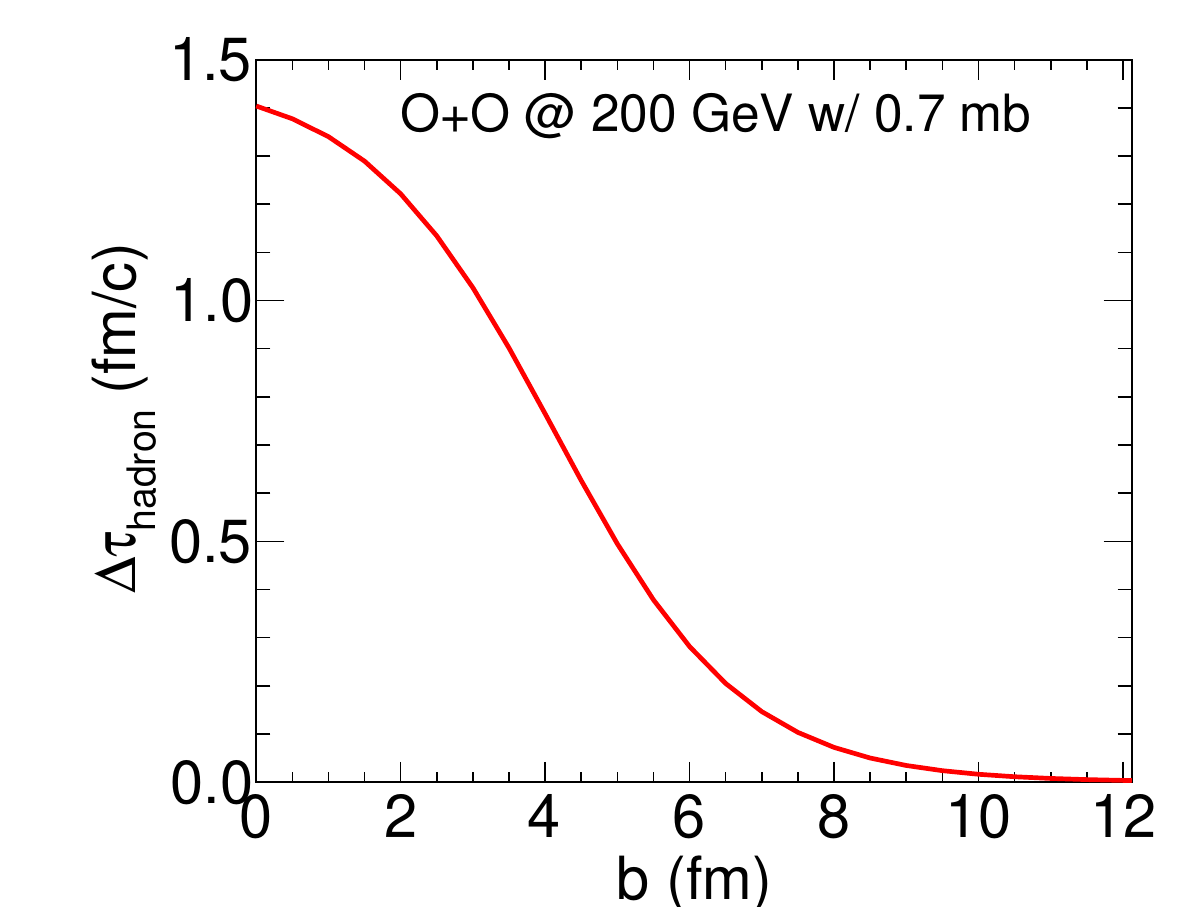}}
\end{minipage}
\caption{The extra hadron proper formation time versus the impact parameter in the AMPT model with parton cross section 0.7 mb.}
\label{fig:en2}
\end{figure}

\begin{figure}[h]
\begin{minipage}[t]{0.95\linewidth}
\subfigure{\includegraphics[width=0.9\textwidth]{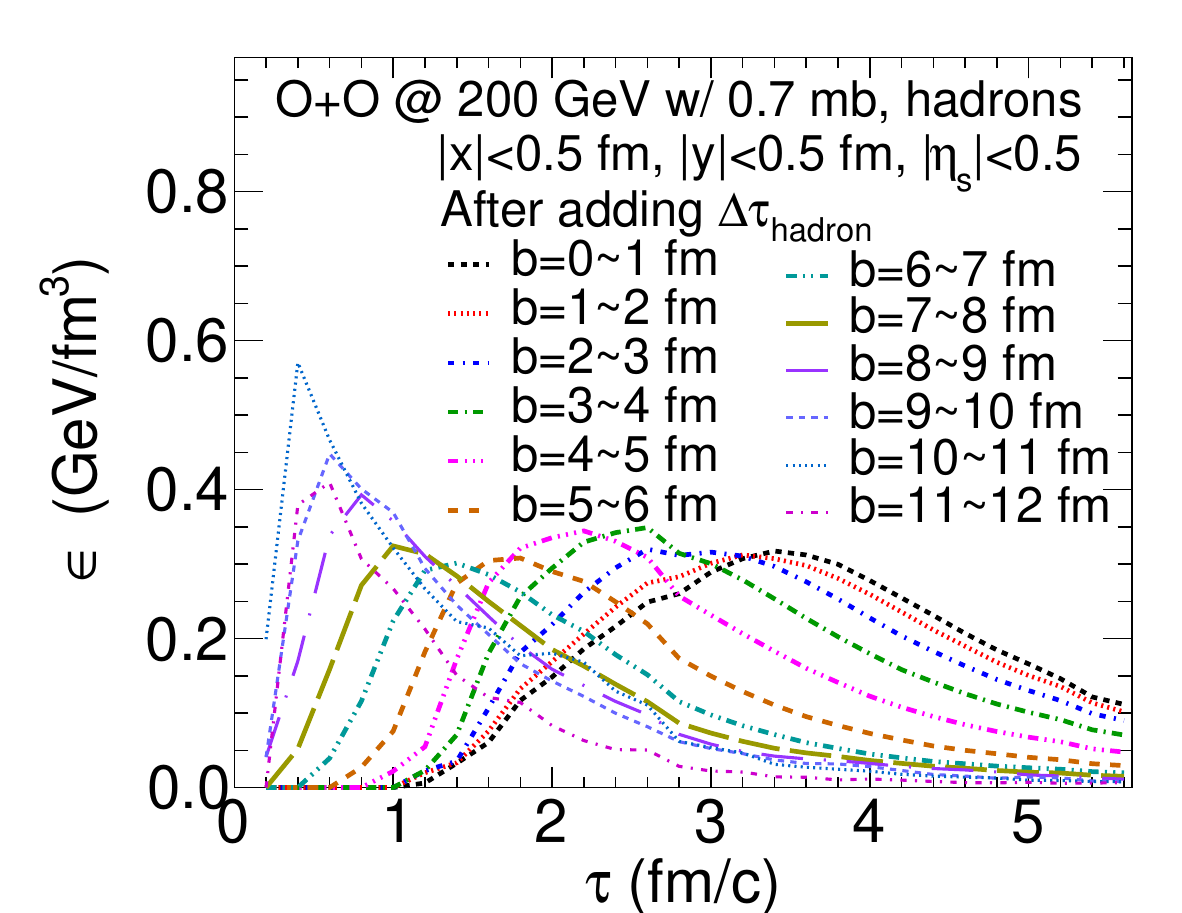}}
\end{minipage}
\caption{The evolution of energy density of hadrons in the center cell of the AMPT model with parton cross section 0.7 mb after adding the extra hadron formation time.}
\label{fig:en3}
\end{figure}

\begin{figure}
\begin{minipage}[t]{0.95\linewidth}
\subfigure{\includegraphics[width=1.0\textwidth]{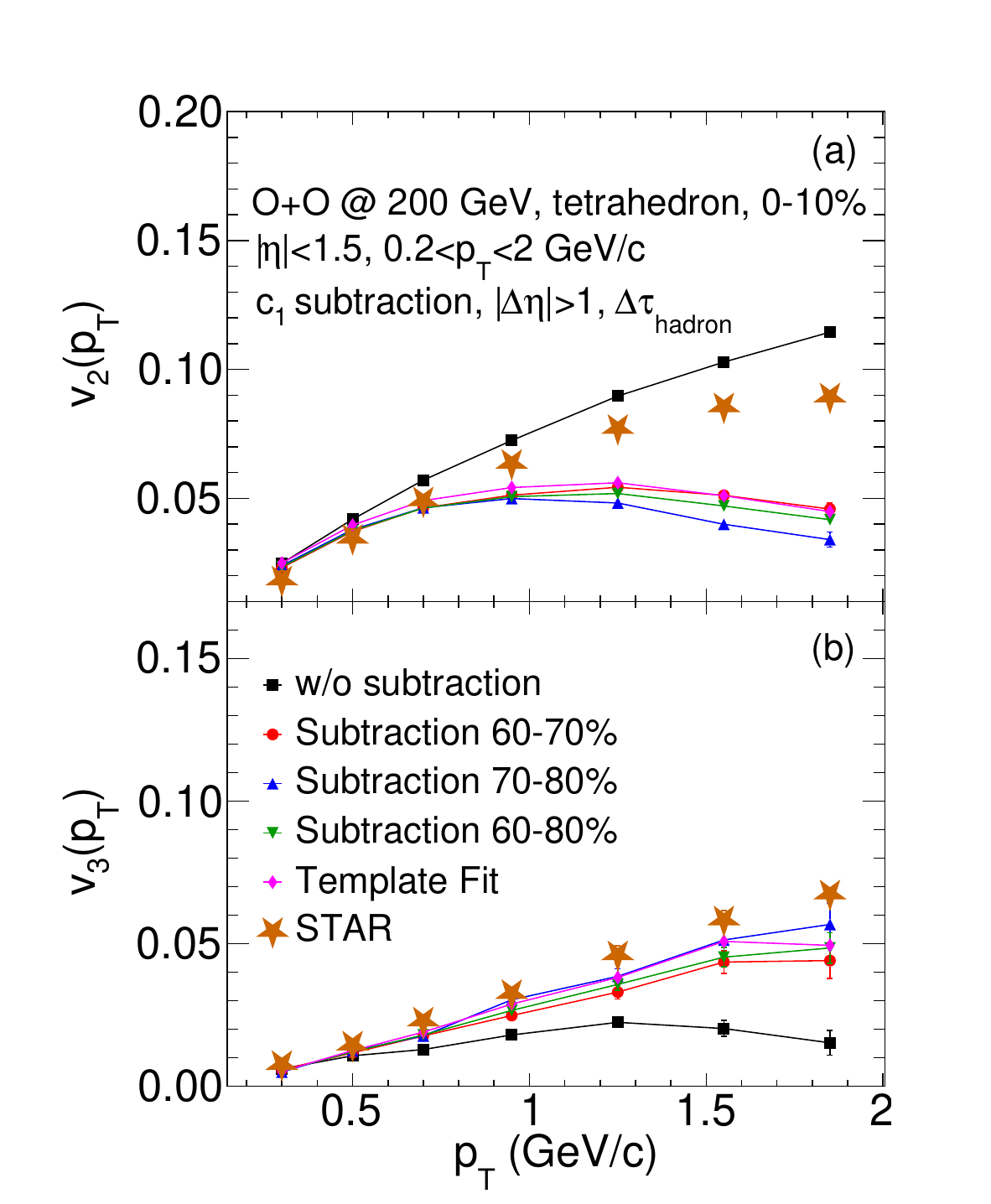}}
\end{minipage}
    \caption{The $p_{\rm T}$ dependence of (a) $v_{2}\{2\}$ and (b) $v_{3}\{2\}$ for 0-10\% centrality bins with different non-flow subtractions in O+O collisions at 200 GeV with tetrahedron configuration, compared to STAR data.}
\label{fig:subdiff}
\end{figure}

\begin{figure*}[htb]
  \begin{minipage}[t]{0.33\linewidth}
    \centering
    \includegraphics[width=0.99\textwidth]{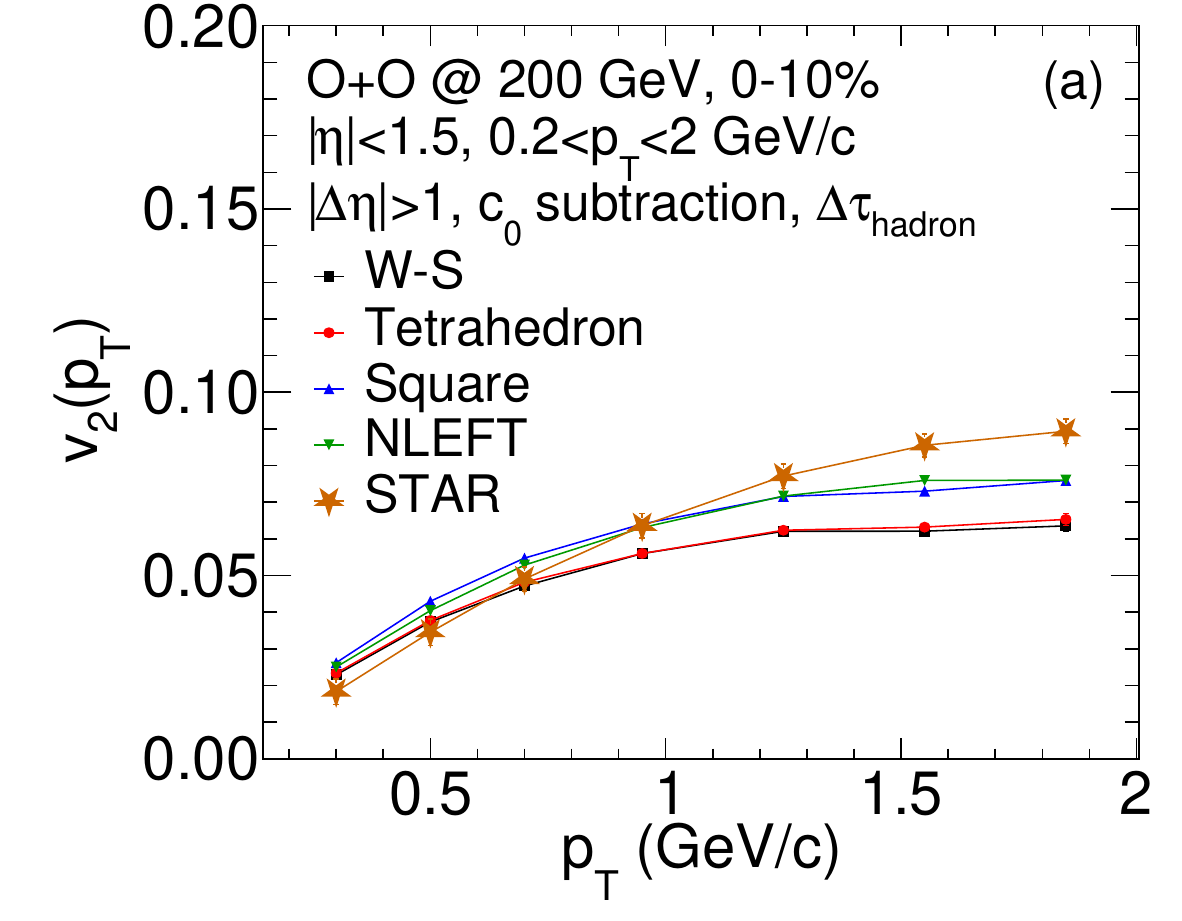}
  \end{minipage}%
     \begin{minipage}[t]{0.33\linewidth}
    \centering
    \includegraphics[width=0.99\textwidth]{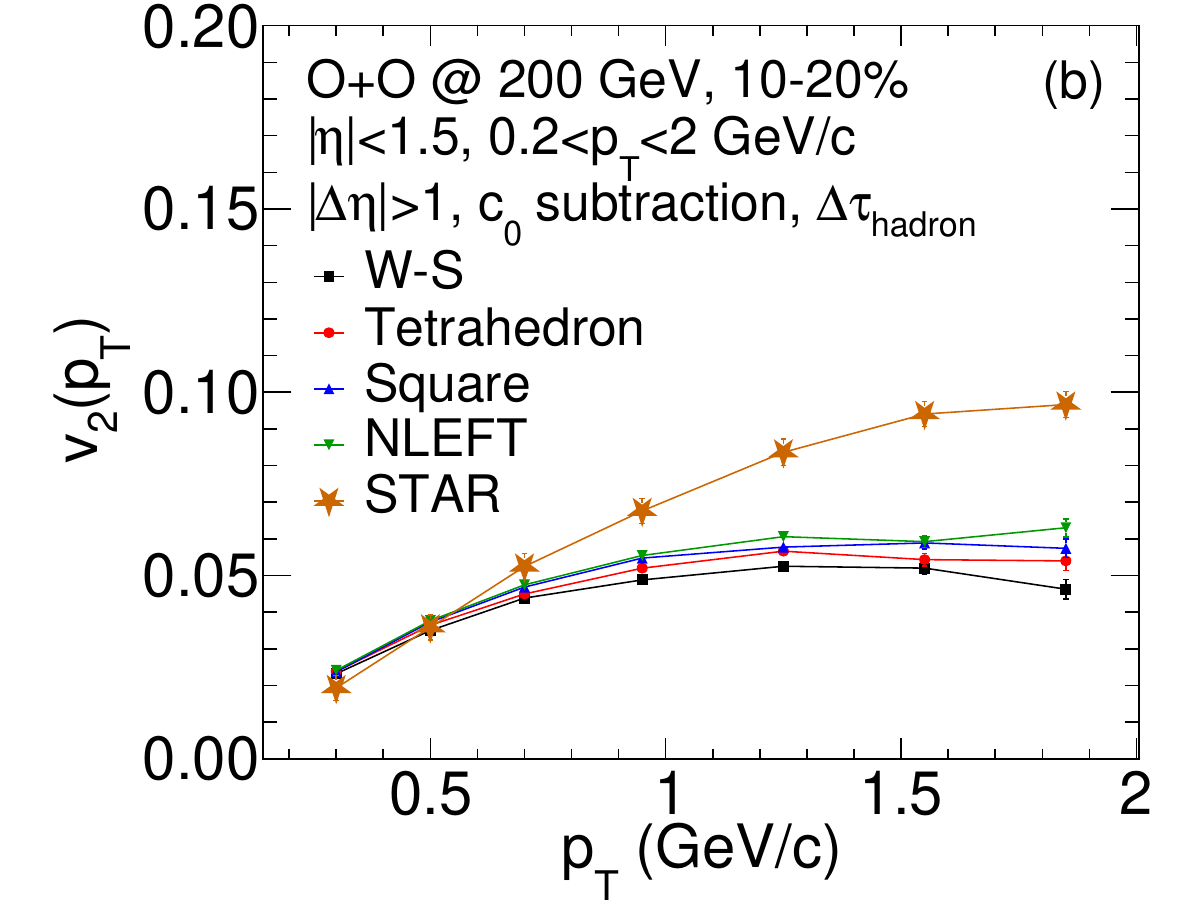}
     \end{minipage}%
     \begin{minipage}[t]{0.33\linewidth}
    \centering
    \includegraphics[width=0.99\textwidth]{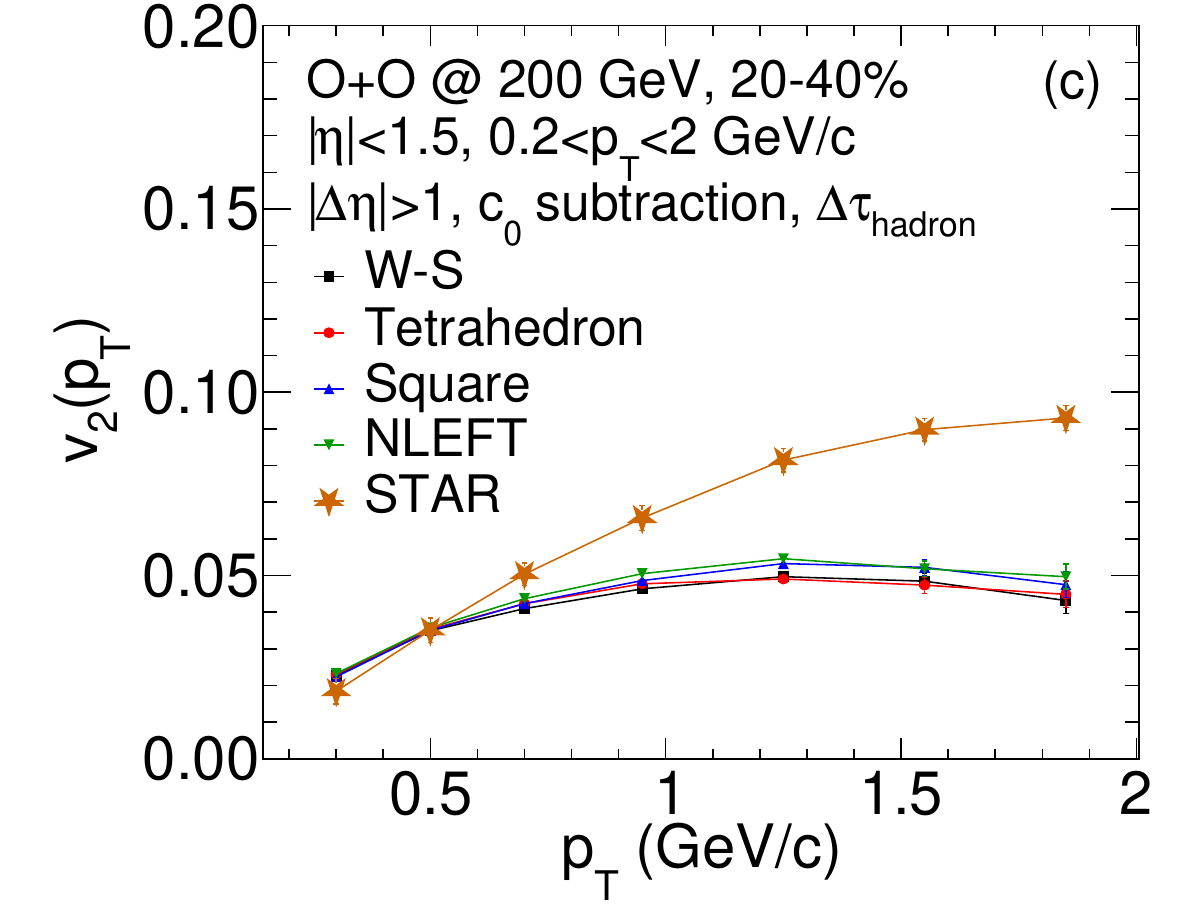} 
  \end{minipage}
  \caption{The $p_{\rm T}$ dependence of $v_{2}\{2\}$ for (a) 0-10\%, (b) 10-20\%, and (c) 20-40\% centrality bins in O+O collisions at 200 GeV with different nuclear structure configurations, compared to STAR data.}
\label{fig:v2pt-c0}
\end{figure*}

\begin{figure*}[htb]
  \begin{minipage}[t]{0.33\linewidth}
    \centering
    \includegraphics[width=0.99\textwidth]{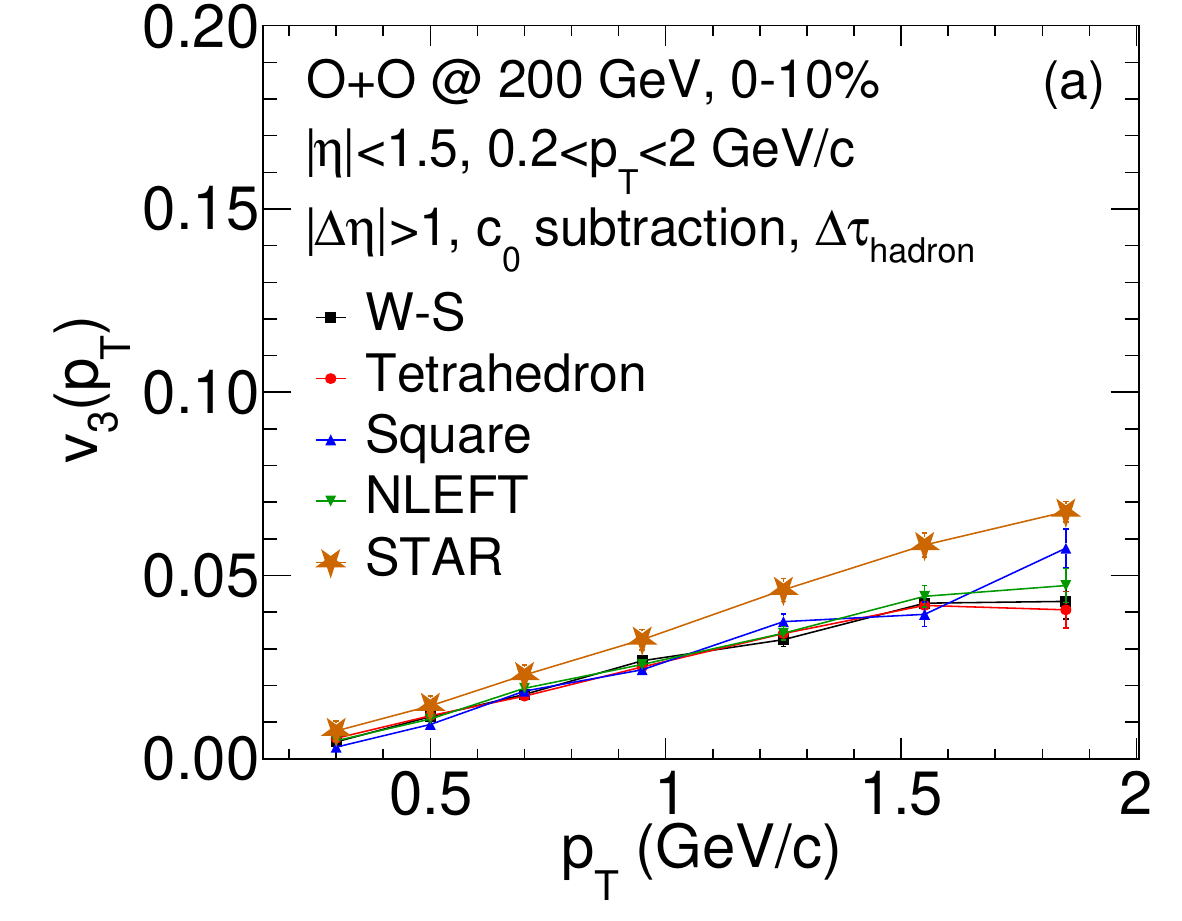}
  \end{minipage}%
     \begin{minipage}[t]{0.33\linewidth}
    \centering
    \includegraphics[width=0.99\textwidth]{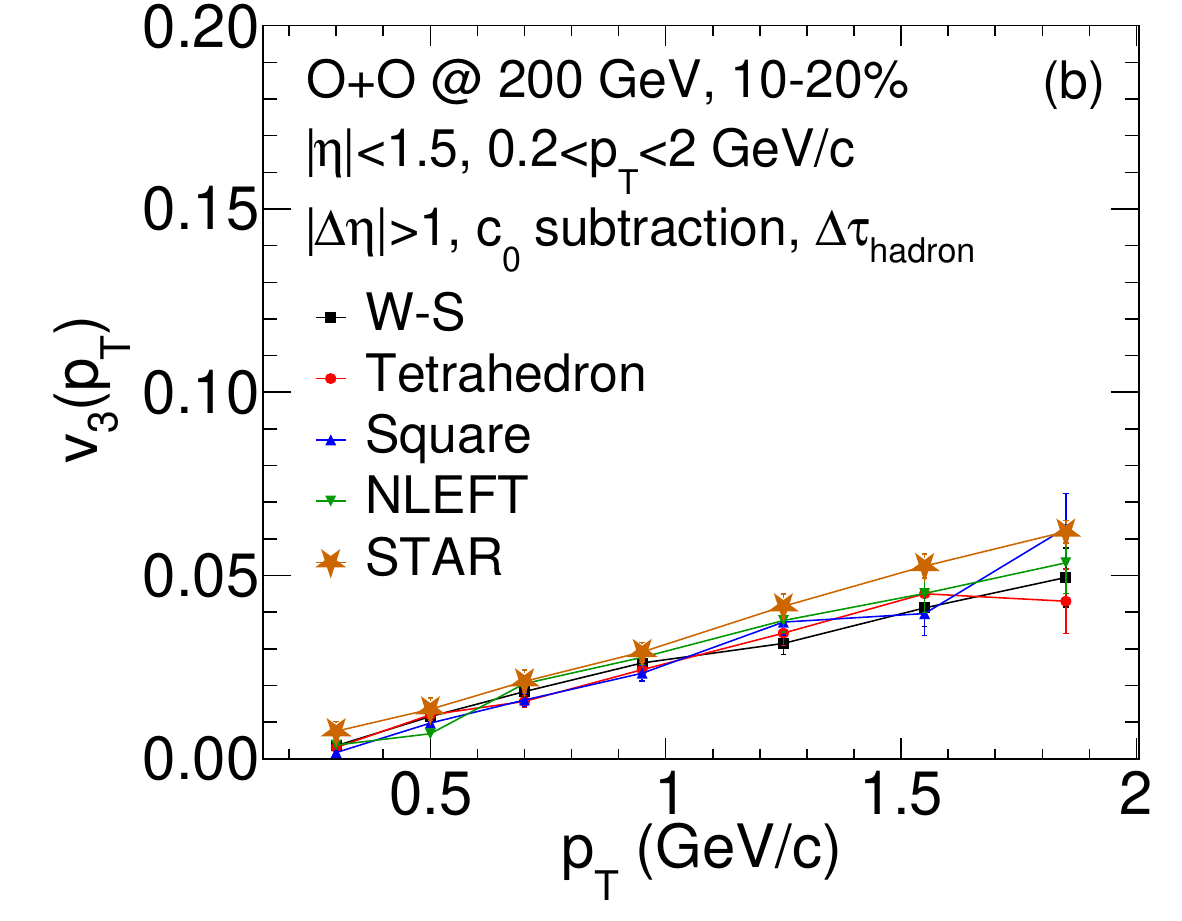}
     \end{minipage}%
     \begin{minipage}[t]{0.33\linewidth}
    \centering
    \includegraphics[width=0.99\textwidth]{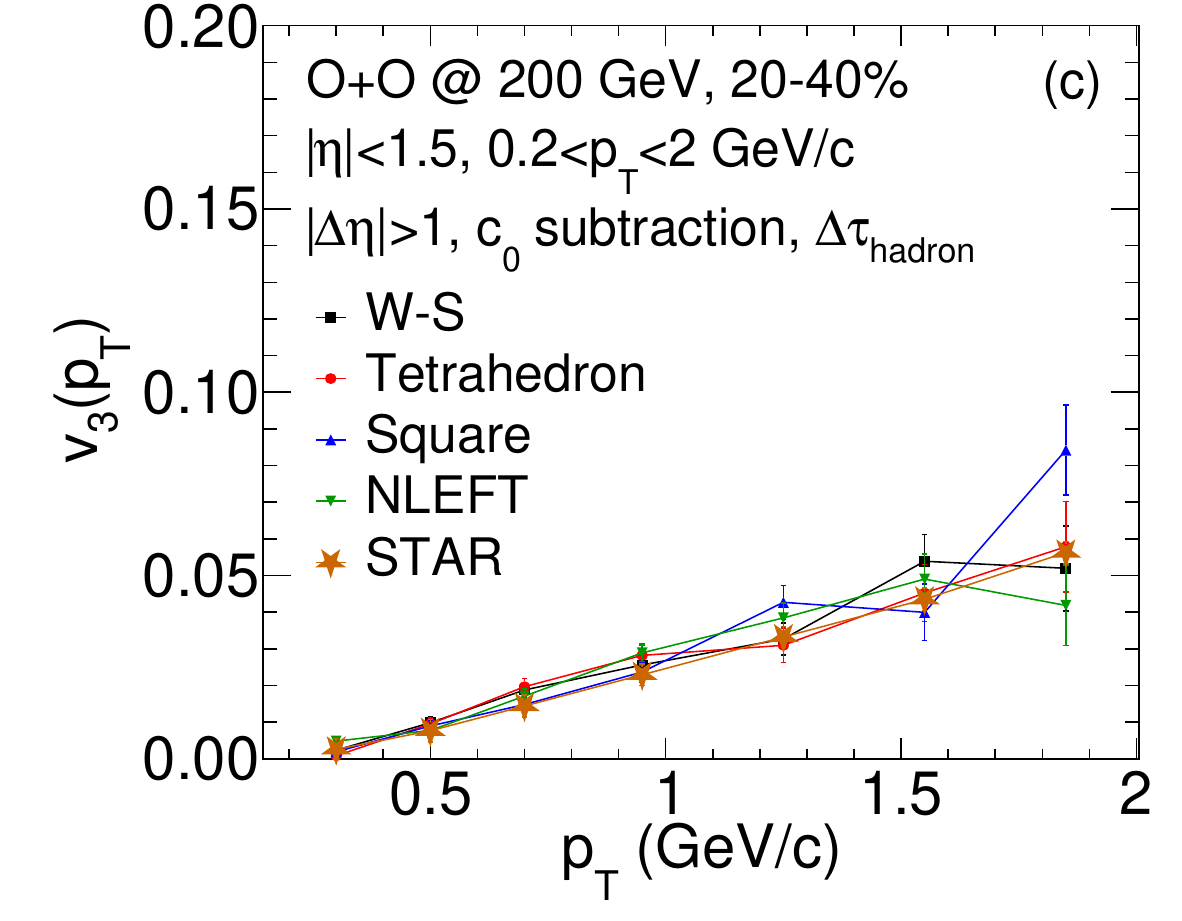} 
  \end{minipage}
    \caption{The $p_{\rm T}$ dependence of $v_{3}\{2\}$ for (a) 0-10\%, (b) 10-20\%, and (c) 20-40\% centrality bins in O+O collisions at 200 GeV with different nuclear structure configurations, compared to STAR data.}
\label{fig:v3pt-c0}
\end{figure*}

\begin{figure}
\begin{minipage}[t]{0.95\linewidth}
\subfigure{\includegraphics[width=1\textwidth]{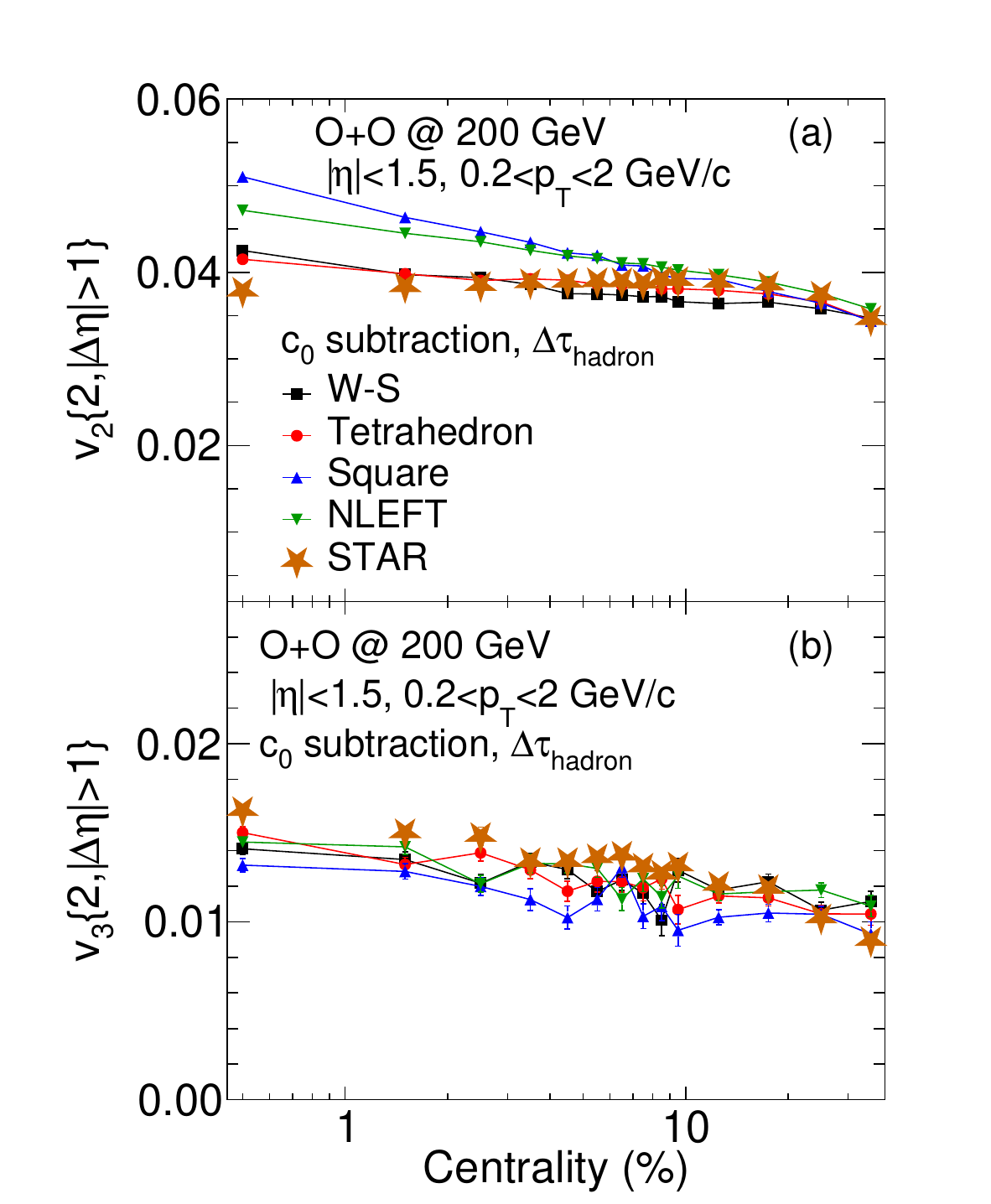}}
\end{minipage}
\caption{The centrality dependence of (a) $v_{2}\{2\}$ and (b) $v_{3}\{2\}$ in O+O collisions at 200 GeV with different nuclear structure configurations, compared to STAR data.}
\label{fig:vn-c0}
\end{figure}

\begin{figure}[htb]
  \begin{minipage}[t]{0.95\linewidth}
    \centering
    \includegraphics[width=0.9\textwidth]{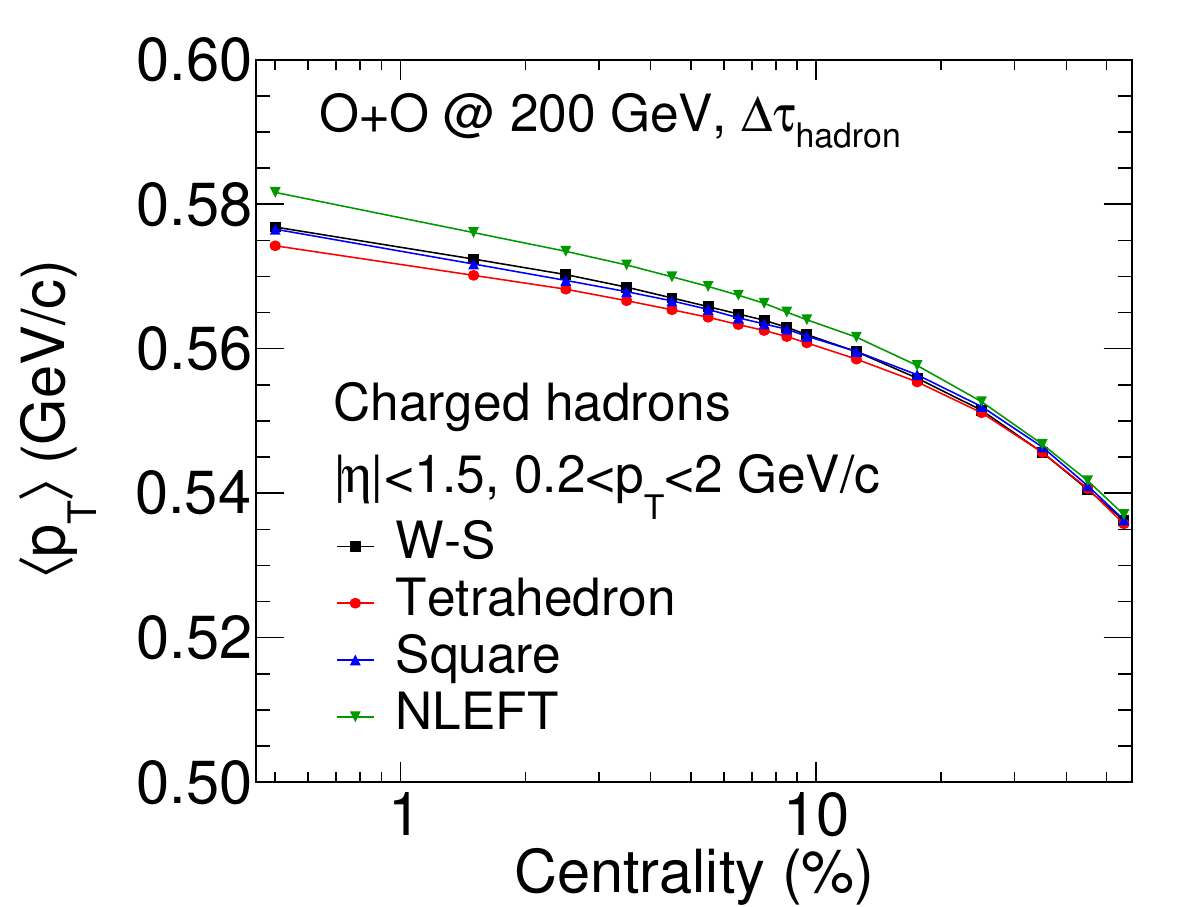}
  \end{minipage}
  \caption{The centrality dependence of $\left<p_{\rm T} \right>$ in O+O collisions at 200 GeV with different nuclear structure configurations.}
\label{fig:mpt1}
\end{figure}

In this study, we have used the AMPT-SM model to conduct a detailed investigation of the cluster structure of the $^{16}$O nucleus in relativistic collisions. We first found that the energy density at the early stage of the hadronic evolution in the AMPT-SM model for O+O collisions is consistently too high for typical values of the parton cross section. Figures~\ref{fig:en0}-\ref{fig:en1} show the evolution of the energy density in the hadron phase for the center cell of O+O collisions for parton cross sections of 1.5 mb and 0.7 mb, respectively, where we see that the maximum energy density for central collisions well exceeds the energy density expected for the QGP phase transition ($\epsilon_{c} \sim 0.3 - 0.8$ GeV/fm$^{3}$). This excess comes from the early freezeout of ZPC for small systems, which then leads to early hadronization and high energy density of the hadronic phase, because quark coalescence starts when parton scatterings in ZPC are finished in the current AMPT-SM model. 
Qualitatively, the density at which ZPC freezes out can be estimated from $n \sigma l \sim 1$, where $n$ is the parton number density, $\sigma$ the parton cross section, and $l$ the system size. Thus at fixed $\sigma$, $n$ is higher for smaller systems where $l$ is smaller. 
For Au+Au collisions, the peak energy density of the hadronic phase for central to mid-central collisions is found to be not far from 0.3 GeV/fm$^{3}$, but it is much higher for O+O collisions (at the same parton cross section). As a result, the early start of the hadronic phase would lead to too many scatterings in the hadronic phase of O+O collisions.

To address this issue, we modified the AMPT-SM model so that the peak energy density in the hadron phase for the center cell at each centrality is around an expected value, chosen to be 0.3 GeV/fm$^{3}$ in this study. 
We implement this modification by increasing the default hadron proper formation time by $\Delta\tau_{\rm hadron}$ that depends on the impact parameter, as illustrated in Fig.~\ref{fig:en2} for the parton cross section of 0.7 mb. For central O+O collisions, an extra formation time of more than 1 fm/$c$ is necessary, while for very peripheral events no extra formation time is needed. As shown in Fig.~\ref{fig:en3}, the added hadron formation time leads to peak energy densities around 0.3 GeV/fm³ for the hadron phase in O+O collisions for almost all centralities. This modified AMPT-SM model is more physical as it decouples the parton cross section from the energy density at hadronization. 
A better way would be to start the quark coalescence dynamically when the local partonic energy density decreases to an expected value; this is left to future work. Note that $\Delta\tau_{\rm hadron}$ is a proper extra formation time, and the formation time in the center-of-mass frame is longer by a $\cosh(y)$ factor with $y$ being the hadron rapidity. Also note that we have not included heavy flavor hadrons in this modification; we have checked that this has little effect on the results of this study.

\section{\label{sec:diffnonflow}Different methods for non-flow subtraction}%

Figures~\ref{fig:subdiff}(a) presents the $v_{2}(p_{\rm T})$ results for the tetrahedron configuration in $0$--$10\%$ centrality, obtained with five different nonflow subtraction methods: no subtraction, subtraction using 60-70\% peripheral events, subtraction with 70-80\% collisions, subtraction with $60-80\%$ collisions, and the template-fit method. In the template-fit method~\cite{ATLAS:2015hzw}, the two-particle correlation function from a peripheral reference sample serves as a non-flow template. After fitting this template to the correlation distribution in central and mid-central events with an additional flow term, the collective flow signal can be extracted by subtracting the non-flow contributions. For comparison, the results obtained directly from Fourier fits without any non-flow subtraction are also shown. As expected, the $v_{2}(p_{\rm T})$ values without subtraction are the largest, particularly at high $p_{\rm T}$, where non-flow contributions are most significant. The $v_{2}(p_{\rm T})$ for non-flow subtractions with 70-80\% collisions is the smallest, possibly because non-flow is over-subtracted. The template-fit method yields similar $v_{2}(p_{\rm T})$ results as non-flow subtractions, with 60-70\% of collisions. Considering that the peripheral subtraction method treats the entire correlation observed in peripheral collisions as non-flow, it typically serves as the lower limit of flow. In contrast, the template-fit addresses the flow modulation in peripheral collisions and thus typically provides an upper limit to the flow.   
Thus, the true flow is expected to lie between the template-fit and peripheral subtraction results. However, the band spanned by these two in Fig.~\ref{fig:subdiff} (a) does not cover the STAR measurement.
Figure~\ref{fig:subdiff} (b) shows the $v_{3}(p_{\rm T})$ with the four different non-flow subtractions and the results without non-flow subtraction. $v_{2}\{2\}$ originates from the initial geometry of the collision region, while $v_{3}\{2\}$ arises from the fluctuations in the initial geometry and subsequent evolution. In contrast to $v_{2}(p_{\rm T})$, the non-flow subtraction approaches have the opposite effect on $v_{3}(p_{\rm T})$: the $v_{3}(p_{\rm T})$ without subtraction is the smallest, while $v_{3}(p_{\rm T})$ for non-flow subtractions with 70-80\% collisions is the largest. We also note that various non-flow subtraction methods do not yield a significant difference in $v_{3}(p_{\rm T})$, which contrasts with the $v_{2}(p_{\rm T})$ case.
The reasonable agreement of AMPT calculations and STAR measurement in $v_{3}(p_{\rm T})$ persists, independent of what non-flow subtraction method is applied. Overall, a more systematic comparison of non-flow subtraction methods between the AMPT model and experimental analyses will be an important direction for future work.

In the main text, we present the anisotropic flow results extracted using the $c_1$ method, which aligns more directly with the experimental approach adopted by the STAR collaboration, as introduced in Sec.~\ref{sec:obs}. For completeness, we provide here the $v_{2}\{2\}$ and $v_{3}\{2\}$ results obtained using the $c_0$ method, including their centrality dependence and transverse momentum ($p_{\rm T}$) dependence. These comparisons help quantify the systematic uncertainty associated with non-flow subtraction procedures in theoretical models.

In the STAR experiment, it has been observed that the $c_{0}$ and $c_{1}$ methods yield nearly identical results for $v_{2}\{2\}$ and $v_{3}\{2\}$ in O+O collisions at $\sqrt{s_{_{\rm NN}}} = 200$ GeV. 
In contrast, our AMPT model calculations reveal non-negligible differences between the two approaches. 
Specifically, the $v_{n}$ values obtained with the $c_{0}$ method are systematically larger than those from the $c_{1}$ method, as shown in Figs.~\ref{fig:v2pt-c0}-\ref{fig:vn-c0}. Figures~\ref{fig:v2pt-c0} and \ref{fig:v3pt-c0} present the $p_{\rm T}$ dependence of $v_{2}\{2\}$ and $v_{3}\{2\}$ from the $c_{0}$ method, while Fig.~\ref{fig:vn-c0} illustrates their centrality dependence. 
The observed discrepancies may arise from model-specific features in the underlying correlation structure or particle production mechanisms, and therefore warrant further theoretical investigation. 
In particular, substantial differences between the $c_{0}$ and $c_{1}$ methods are evident in the $p_{\rm T}$-differential anisotropic flow in peripheral collisions and at high $p_{\rm T}$. 
This underscores the importance of carefully understanding and constraining non-flow subtraction methods when interpreting flow signals in small systems. 
It is also worth noting that the ordering of the four configurations remains consistent between the $c_{0}$ and $c_{1}$ methods.

\section{\label{sec:mpt} $\left<p_{\rm T} \right>$ results for O+O collisions}%

Figure~\ref{fig:mpt1} presents the centrality dependence of $\langle p_{\rm T}\rangle$ of charged hadrons in O+O collisions under STAR-matched kinematics $|\eta|<1.5$ and $0.2<p_{\rm T}<2$~GeV/$c$. The results are presented for four $^{16}$O configurations: W-S, tetrahedron, square and NLEFT. 
The differences in $\left<p_{\rm T} \right>$ across configurations are pronounced, especially between the NLEFT and tetrahedron cases. For central and mid-central collisions, the $\left<p_{\rm T} \right>$ from NLEFT is higher than the other configurations, while the $\left<p_{\rm T} \right>$ from tetrahedron configuration is the smallest. The $\left<p_{\rm T} \right>$ values for the W-S and square configurations are similar. Given the considerable sensitivity of $\left<p_{\rm T} \right>$ to $v_{2}\{2\}$, it is reasonable to expect a similar sensitivity of $v_{2}\{2\}$ to different geometric configurations.

\end{appendices}

\end{document}